\documentclass[a4paper,11pt]{article}
\pdfoutput=1 

\usepackage{jcappub}
\usepackage{slashed}

\usepackage[T1]{fontenc} 
\usepackage{caption}
\usepackage{subcaption}

\title{\boldmath Randers geometry as MOND/dark matter}

\author{Qasem Exirifard}

\affiliation{
Institute for Research in Fundamental Sciences (IPM),
\\Tehran, Iran
}

\emailAdd{exir@theory.ipm.ac.ir}

\abstract{
We consider a deviation of the physical length  from the Riemann geometry toward the Randers'. We construct  a consistent second-order  relativistic theory of gravity that dynamically reduces to the Einstein-Hilbert theory for the strong and Newtonian gravity while its weak gravitational regime reproduces MOND and the gravitational lensing attributed to the dark matter halo. It also naturally accommodates  the observed value of the cosmological constant. We show that it predicts a few percent deviation for the post Newtonian parameter $\gamma$ in a part of the regime that interpolates the Newtonian regime to the MOND regime.  The deviation  is consistent with the reported observations but can possibly be detected by fine-tuned refinements of the current data or specified future observations. }

\begin{document}

\maketitle
\flushbottom

\section{Introduction}
 MOND was proposed in 1983 by Mordehai Milgrom \cite{MOND}. It was, then, altered to the AQUAL theory \cite{AQUAL},  a non-covariant modified theory  of  gravity capable of reproducing MOND.  TeVeS, as a relativistic MOND theory, was proposed  by Bekenetsein in 2004   \cite{Bekenstein:2004ne}. TeVeS theory, despite its successes, does not reduce to the Einstein-Hilbert gravity for strong gravity. The purpose of this research is to introduce a relativistic MOND theory that reduces to the Einstein-Hilbert gravity in the strong regimes of gravity.
In so doing we introduce a deviation of the Riemann  geometry toward Finsler/Randers geometry as the cause of dark paradigms.

 We adapt the Randers geometry as the physical geometry in the section \ref{RandersSection}. The dynamical degrees of the freedom in the adapted geometry are scalars, gauge fields and a metric. We introduce two gauge fields and two scalars. This introduction requires four coupling constants, a functional and a length scale.

 In the section \ref{dynamics} we fix the dynamics such that the strong gravity limit of the theory coincides to that of the Einstein-Hilbert gravity.  This reduces the number of coupling constants  to two. This causes one of the gauge fields and one of the scalars to have the kinetic energy  of ghost. We prove that the classical solutions coincides to those of the Einstein-Hilbert theory.  We also prove that the ghosts are decoupled from the physical spectrum by performing an exact integration over the gauge fields and scalars in the Feynman path integral. Notice that   this does not constrain the dynamics of the fields in the weak regime of gravity/fields. We shall choose the dynamics such that the theory reproduces MOND.

 The section \ref{staticsolutions} reports the static solution in the Newtonian and MOND regimes of the theory. The sub-section \ref{OrbitOfParticles} calculates the light trajectory around a static solution. We require the $\gamma$ -the post-Newtonian parameter of theory- to be $1$ in the deep MONDian regime. This requirement reduces the coupling constants of the theory to one constant and maps the length parameter of the theory to the critical acceleration in the MOND.

We consider a family of the interpolating functionals in the section \ref{WeakGravitationalLensing}. We ignore the negligible contribution of the gauge fields and scalars to the metric in the MOND regime. We then find the exact static spherical solution of the theory.  We calculate the $\gamma$ parameter around spherical static solutions in whole of the space including the regime that interpolates the Newtonian regime to the MOND regime. We report that requiring the $\gamma$ parameter in the interpolating regime to be consistent with ref. \cite{Amendola:2013qna} sets a strong upper bound on the coupling constant of the theory as depicted in the fig. \ref{knplotWGL}.  We observe that these bounds  are in agreement with the value of $\gamma$ measured at the surface of the Sun as depicted in fig. \ref{fig:SunGamma}.

We study the Lunar system constraints on the theory in the section \ref{LunarSystemConstraints}. We report that Laser Lunar Ranging measurements \cite{LLR} combined with the LAGEOS data \cite{LAGEOS} provide a lower bound on the coupling constant of the theory, as depicted in fig \ref{fig:Lunar} and \ref{fig:Lunar2} for various values of $p$ that encodes the precision of the used perturbation theory around the Earth.

We study the consistency of the spherical static solution in the section \ref{ConsistencySection}. We see that as we move toward the asymptotic infinity, the neglected contribution of the gauge-fields and scalars to the energy-momentum tensor  accumulates in the deep-MOND regime. For any given mass, we calculate the length scale wherein the energy-momentum tensor of the gauge and scalar fields must be considered, the result is expressed in \eqref{LConsistency}. We report that for any given astronomical mass, this length scale is far larger than the size of the observed Universe. This is one more illustration  that the ghosts do not affect the classical solutions. We do not study how to quantise gravity that here is composed of a metric and two scalars and two gauge fields. We suffice to the study of the classical solutions (IR solutions) and we show that the theory is consistent in its IR limit. Since the strong field limit of the theory coincides to the Einstein-Hilbert theory we expect its UV limit to be the same as the UV limit of the Einstein-Hilbert theory.   

There is a freedom in identifying the constant of the integration in the functional of the theory. We use this freedom and naturally accommodate the observed value of the cosmological constant in the theory in the section of \ref{CosmologiucalConstant} with a tuning at the order one. This accommodation identifies the only free coupling constant of the theory.  The identified theory predicts that the $\gamma$ parameter in the interpolating regime deviates from $1$ by few percents only in a section of the interpolating regime. This prediction can be attested by refining the current data or specified future observations.

We study the cosmology of the theory in the section \ref{Cosmology}. We slightly change the coupling of the gauge fields to the matter current in order to have a homogeneous isotropic time-dependent solution. This change leaves intact the static solutions of the previous sections. We report that the isotropic homogeneous solution of the theory coincides to that of the Einstein-Hilbert theory with a cosmological constant. So the late time cosmology of the theory is in accord with observation.

We discuss on the results and concludes in the section \ref{summary}.

\section{Adapting the Randers geometry}\label{RandersSection}
We assume that the orbit of a particle of mass $m$ is derived from the variation of
\begin{equation}\label{Randers.1}
S[x(\tau)]=- m \int d\tau (\sqrt{ e^{2 \bar{\phi}} g_{\mu\nu} \dot{x}^\mu \dot{x}^\nu} + \bar{A}_\mu \dot{x}^\mu)\,,
\end{equation}
where $\bar{\phi}$ is an scalar, $\bar{A}_\mu$ is a gauge field and $g_{\mu\nu}$ represents the metric. This is the definition of length in the Randers geometry \cite{Randers}. Notice that the flat metric in our notation reads:
\begin{equation}
\eta_{\mu\nu} = (1,-1,-1,-1).
\end{equation}
Now let a positive auxiliary field $\eta(\tau)$ be defined on the world-line of the particle and consider
\begin{equation}\label{Randers.2}
S[x(\tau), \eta(\tau)]=- m \int d\tau
\left(
\frac{1}{2 \eta(\tau)} e^{2 \bar{\phi}} g_{\mu\nu} \dot{x}^\mu \dot{x}^\nu +\frac{\eta(\tau)}{2} + \bar{A}_\mu \dot{x}^\mu
\right) \,,
\end{equation}
the variation of which with respect to $\eta(\tau)$ yields
\begin{equation}
\frac{\delta S}{\delta \eta} = 0 \to \eta(\tau) = e^{\bar{\phi}}\sqrt{ g_{\mu\nu} \dot{x}^\mu \dot{x}^\nu}\,,
\end{equation}
inserting which into \eqref{Randers.2} yields \eqref{Randers.1}. So  \eqref{Randers.2} and \eqref{Randers.1} are equivalent.  \eqref{Randers.2} is invariant under the reparameterization of the world-line:
\begin{eqnarray}
\tau &\to& \tilde{\tau} = \tilde{\tau}(\tau)\,\\
\eta(\tau) &\to& \tilde{\eta}(\tilde{\tau}) = \frac{d \tau}{d\tilde \tau} \eta(\tau).
\end{eqnarray}
This reparameterization allows to set
\begin{equation}
\eta(\tau) =1\,,
\end{equation}
which, hereafter, is referred to as the standard parameterization.   The standard parametrization simplifies \eqref{Randers.2}  to
\begin{equation}\label{Randers.3}
S[x(\tau)]=- m \int d\tau
\left(
\frac{1}{2} e^{2 \bar{\phi}} g_{\mu\nu} \dot{x}^\mu \dot{x}^\nu + \bar{A}_\mu \dot{x}^\mu
\right) \,,
\end{equation}
where
\begin{equation}
\label{gaugefixingparameter}
e^{2 \bar{\phi}} g_{\mu\nu} \dot{x}^\mu \dot{x}^\nu = 1\,,
\end{equation}
is understood as the gauge-fixing constraint imposed on its equations of motion. Notice that \eqref{Randers.3} is equivalent to \eqref{Randers.1} but it is simpler. We will make extensive use of this simpler Lagrangian througout this work.\footnote{To string theorists, the difference between \eqref{Randers.3} and \eqref{Randers.1}  is similar to the difference between Numbo-Goto action and Polyakov action. We could have worked with \eqref{Randers.1} but  in analogy with the section $2.1$ of   \href{http://www.blau.itp.unibe.ch/newlecturesGR.pdf}{the   Matthias Blau's lecture notes on GR}, we have preferred to work with the simpler Lagrangian.} 

\section{Dynamics}
\label{dynamics}
We wish to construct the dynamics of the gauge field $\bar{A}_\mu$ and $\bar{\phi}$ such that the theory admits two regimes:
\begin{enumerate}
\item \textbf{EH regime}: strong gravity reproduces the Einstein-Hilbert gravity,
\item \textbf{MOND}: very weak gravity reproduces MOND and gravitational lensing attributed to a dark matter.
\end{enumerate}
To ensure the existence of the EH regime, we suggest that $\bar{A}_\mu$ and $\bar{\phi}$ are composite,
\begin{eqnarray}
\bar{A}_\mu &=& A_1 + A_2\,,\\
\bar{\phi} & =& \phi_1 + \phi_2\,.
\end{eqnarray}
We assume that the strong regime of gravity  holds
\begin{subequations}
\label{EHregime0}
\begin{eqnarray}
\bar{A}_\mu &=& d_\mu \Lambda\,,\\
\bar{\phi} &=& \text{cte},
\end{eqnarray}
\end{subequations}
So the orbits of particles derived from \eqref{Randers.3} in the EH regime  are identical to those derived from
\begin{equation}
S[x(\tau)] =- \frac{1}{2} m e^{2\bar{\phi}} \int d\tau g_{\mu\nu} \dot{x}^\mu \dot{x}^\nu\,,
\end{equation}
which is the effective action of a particle with the effective mass of $m e^{2\phi}$ in the Einstein-Hilbert gravity. Due to the gauge symmetry of the theory and the freedom in the mass definition, we choose the following representation  for \eqref{EHregime0}:
\begin{subequations}
\label{EHregime}
\begin{eqnarray}
\bar{A}_\mu &=& 0,\\
\bar{\phi} &=& 0.
\end{eqnarray}
\end{subequations}
We assume that the dynamics of the metric is given by the Einstein-Hilbert action:
\begin{equation}
S[g] \equiv \frac{1}{16 \pi G } \int d^4 x \sqrt{-\det g} R\,.
\end{equation}
In order to write the matters' action we consider a set of particles $m_i, i \in  \{1, N\}$.   The world-line of these particles can be parameterized by a global choice of time in a smooth space-time geometry. The cumulative action of these particles, followed from \eqref{Randers.3}, reads
\begin{equation}
\label{massterdiscontinuus}
S_M = \int dt
\left(
-\sum_{i=1}^N \frac{1}{2} m_i (e^{2\bar{\phi}} g_{\mu\nu} \dot{x}^\mu \dot{x}^\nu-1) - \sum_{i=1}^N m_i \bar{A}_\mu \dot{x}^\mu
\right) \,,
\end{equation}
where the non-gravitational interactions between the particles are neglected for sake of simplicity. Also notice that an appropriate constant term has been added to the action in order to get the energy-momentum tensor of pressure-less fluid in the continuum limit \eqref{EnergyMomentumTensorMatter}.  In the continuum approximation, \eqref{massterdiscontinuus} is mapped to
\begin{equation}
\label{MatterAction}
S_{M}[g_{\mu\nu},  \bar{A}_\mu, \bar{\phi}, u^\mu,\rho] = \int d^4 x \sqrt{-\det g} \left(
-\frac{1}{2} \rho( e^{2 \bar{\phi} } g_{\mu\nu} u^\mu u^\nu-1) - \bar{A}_\mu \rho u^\mu
\right)\,,
\end{equation}
where $\rho$ is the density and $u^\mu$ is the four  vector current of the matter distribution. Notice that due to \eqref{gaugefixingparameter}, it holds
\begin{equation}
\label{umuconst}
e^{2\bar{\phi}} g_{\mu\nu} u^\mu u^\nu = 1\,,
\end{equation}
which should be imposed at the level of the equations of motion. 

We set the action of the gauge fields to
\begin{subequations}
\label{ActionGauges}
\begin{eqnarray}
S[A_1]&=& - \frac{1}{16 \pi \kappa_1 G} \int d^4x \sqrt{-\det g} \,\frac{1}{4} F^1_{\mu\nu} F_{1}^{\mu\nu}\,,\\
S[A_2]&=& \frac{1}{16 \pi \kappa_2 G l^2}\int d^4x \sqrt{-\det g}\, {\cal L}(-\frac{l^2}{4}F^2_{\mu\nu} F_{2}^{\mu\nu}),
\end{eqnarray}
\end{subequations}
where $F^1_{\mu\nu}$ and $F^2_{\mu\nu}$ are the gauge field strengths:
\begin{subequations}
\begin{eqnarray}
F^1_{\mu\nu}&=& \partial_\mu A^1_{\nu} - \partial_\nu A^1_{\mu}\,,\\
F^2_{\mu\nu}&=& \partial_\mu A^2_{\nu} - \partial_\nu A^2_{\mu}\,,
\end{eqnarray}
\end{subequations}
and $l$ is a parameter, $\kappa_1$  and $\kappa_2$ are the coupling constants, and $\cal L$ is a functional with the following asymptotic behaviors:
\begin{equation}
\label{Ldefition}
{\cal L}'(x) \equiv \frac{d {\cal L}}{dx} =
\left\{
\begin{array}{cccr}
1 &,& x \gg 1 & \text{EH regime}\\
x^{\frac{1}{2}} &,& 0\le x \le 1 & \text{~~~~~~~MOND regime}
\end{array}
\right.\,,
\end{equation}
where $x>0$ is understood. 
We shall see that $x\gg1$ represents the EH regime while $x\le 1$ represents the MOND regime of the theory.
We set the action of the scalars to
\begin{subequations}
\begin{eqnarray}
S[\phi_1]&=& - \frac{1}{16 \pi \tilde{\kappa}_1 G} \int d^4x \sqrt{-\det g} \,\frac{1}{2} \partial_\mu \phi_1 \partial^\mu \phi_1\,,\\
S[\phi_2]&=& \frac{1}{16 \pi \tilde{\kappa}_2 G l^2}\int d^4x \sqrt{-\det g}\, {\cal L}(-\frac{l^2}{2 } \partial_\mu \phi_2 \partial^\mu \phi_2),
\end{eqnarray}
\end{subequations}
where $\tilde{\kappa}_1$ and $\tilde{\kappa}_2$ are the coupling constants,  $l$  and $\cal L$ are introduced in the action of the gauge fields.

The equations of motion of the gauge fields are derived from the variation of \eqref{MatterAction} and  \eqref{ActionGauges} :
\begin{subequations}
\label{gaugeequations}
\begin{eqnarray}
\nabla_\nu F^{\nu\mu}_1 & =& 16 \pi \kappa_1 G J^\mu\,,\\
\label{gaugeequationsb}
\nabla_\nu \left({\cal L}'(-\frac{l^2}{4 }F^2_{\alpha\beta} F_{2}^{\alpha\beta})F^{\nu\mu}_2\right) & =& 16 \pi \kappa_2 G J^\mu\,.
\end{eqnarray}
\end{subequations}
where $J^\mu= \rho u^\mu$ and ${\cal L}'(z)= \frac{d {\cal L}(z)}{d z}$.
The equations of the scalars follow:
\begin{subequations}
\label{ScalarsEquations}
\begin{eqnarray}
\Box \phi_1 &=& 16 \pi \tilde{\kappa}_1 G \rho \,,\\
\label{ScalarsEquationsb}
\nabla_\nu \left({\cal L}'(-\frac{l^2}{2 } \partial_\mu \phi_2 \partial^\mu \phi_2) \nabla^\nu\phi_2 \right) &=& 16 \pi \tilde{\kappa}_2 G \rho \,,
\end{eqnarray}
\end{subequations}
where \eqref{umuconst} is utilized.

In order to write the equation of the metric we first compute the energy-momentum tensor:
\begin{eqnarray}
\label{EnergyMomentumTensor}
T_{\mu\nu} & =& T_{M\mu\nu} + \sum_{\alpha=1}^2 T_{A_{\alpha} \mu\nu} + \sum_{\alpha=1}^2 T_{\phi_{\alpha} \mu\nu} ,\\
T_{M\mu\nu} &=& -\frac{2}{\sqrt{-\det g}} \frac{\delta S_M}{\delta g^{\mu\nu}}= \rho e^{2\bar{\phi}} u_\mu u_\nu\,,
\label{EnergyMomentumTensorMatter}
\\
T_{A_1\mu\nu} &=& -\frac{2}{\sqrt{-\det g}} \frac{\delta S[A_1]}{\delta g^{\mu\nu}}\,\,=
\, \frac{1}{16\pi \kappa_1 G}
\left(
F_{1\mu}^{~\alpha} F_{1\nu\alpha} - \frac{1}{4}  g_{\mu\nu} F_{1}^{~\alpha\beta} F_{1\alpha\beta}
\right)\,,
\\
T_{A_2\mu\nu} &=& -\frac{2}{\sqrt{-\det g}} \frac{\delta S[A_2]}{\delta g^{\mu\nu}}\,=
\, \frac{1}{16\pi \kappa_2 G}
\left(
{\cal L}'(l^2 \frac{||F_{2}||}{4}) F_{2\mu}^{~\alpha} F_{2\nu\alpha} - \frac{1}{l^2} g_{\mu\nu}  {\cal L}(l^2 \frac{|| F_{2}|| }{4})
\right),\nonumber\\
\\
T_{\phi_1\mu\nu} &=& -\frac{2}{\sqrt{-\det g}} \frac{\delta S[\phi_1]}{\delta g^{\mu\nu}}\,= \frac{1}{16 \pi \tilde{\kappa}_1 G } \left(
\partial_\mu \phi_1 \partial_\nu \phi_1 -\frac{1}{2} |\partial \phi_1|^2 g_{\mu\nu}
\right)\,,\\
T_{\phi_2\mu\nu} &=& -\frac{2}{\sqrt{-\det g}} \frac{\delta S[\phi_2]}{\delta g^{\mu\nu}}=
\frac{1}{16 \pi \tilde{\kappa}_2 G } \left( {\cal L}'(l^2 \frac{ |\partial \phi_2|^2}{2 } )
\partial_\mu \phi_2 \partial_\nu \phi_2 - \frac{1}{l^2} g_{\mu\nu} {\cal L}(l^2 \frac{|\partial \phi_2|^2}{2} )
\right)\,,\nonumber\\
\end{eqnarray}
where
\begin{eqnarray}
|\partial \phi_2|^2 &=& -\partial_\mu \phi_2 \partial^\mu \phi_2\,,\\
||F_2|| &=& - F_{2\mu\nu} F_{2}^{\mu\nu}\,.
\end{eqnarray}
Notice that \eqref{umuconst} is utilised to simplify \eqref{EnergyMomentumTensorMatter} and write in the form of the standard energy-momentum tensor of pressure-less fluid:
\begin{equation}
T_{M\mu\nu} = \rho_{eff} u^\nu u^\mu\,,
\end{equation} 
where
\begin{equation}
\rho_{eff} = e^{2\bar{\phi}} \rho \,.
\end{equation} 
The equation of motion of the metric then reads
\begin{equation}
\label{metricGTall}
G_{\mu\nu} = 8 \pi G T_{\mu\nu}\,.
\end{equation}
The Einstein-Hilbert regime of \eqref{Ldefition}  simplifies the equation of motion of the gauge fields \eqref{gaugeequations} and the scalars \eqref{ScalarsEquations} and yields:
\begin{eqnarray}
\nabla_\nu \bar{F}^{\nu\mu} &=& 16 \pi (\kappa_1+\kappa_2) G J^\mu\,,\\
\Box \bar{\phi}&=& 16 \pi  (\tilde{\kappa}_1+\tilde{\kappa}_2) G \rho\,,
\end{eqnarray}
where $\bar{F}_{\nu\mu}= \partial_\mu \bar{A}_\nu - \partial_\nu \bar{A}_\mu$. The consistency with \eqref{EHregime0} demands that
\begin{subequations}
\label{ConsCoupling}
\begin{eqnarray}
\kappa_1+\kappa_2&=&0\,,\\
\tilde{\kappa}_1+\tilde{\kappa}_2 & =& 0\,.
\end{eqnarray}
\end{subequations}
Now consider a gauge field $\mathbf{A}_\mu$ and its field strength $\mathbf{F}_{\mu\nu}$ and scalar $\mathbf{\phi}$ that solve
\begin{eqnarray}
\nabla_\nu \mathbf{F}^{\nu\mu} &=& 16 \pi G J^\mu\,,\\
\Box \mathbf{\Phi} & =& 16 \pi G \rho\,.
\end{eqnarray}
After imposing appropriate boundary conditions on $A_{1}$ and $A_{2}$, and on $\phi_1$ and $\phi_2$, one can writes
\begin{subequations}
\begin{eqnarray}
A_1^\mu & = & \kappa_1 \mathbf{A}^\mu \,,\\
A_2^\mu & = & \kappa_2 \mathbf{A}^\mu  \,,
\end{eqnarray}
and
\begin{eqnarray}
\phi_1 & = & \tilde{\kappa}_1 \mathbf{\Phi} \,,\\
\phi_2 & = & \tilde{\kappa}_2 \mathbf{\Phi}  \,,
\end{eqnarray}
\end{subequations}
utilizing which returns:
\begin{eqnarray}
T_{A_1\mu\nu} + T_{A_2\mu\nu}&=&  (\kappa_1+ \kappa_2) \mathbf{T}_{\mu\nu}^{A} =0\,,\\
T_{\phi_1\mu\nu} + T_{\phi_2\mu\nu}&=&  (\tilde{\kappa}_1+ \tilde{\kappa}_2) \mathbf{T}_{\mu\nu}^{\Phi} =0  \,,
\end{eqnarray}
where
\begin{eqnarray}
\mathbf{T}_{\mu\nu}^{A} &=& \frac{1}{16\pi  G}
\left(
\mathbf{F}_{\mu}^{~\alpha} \mathbf{F}_{\nu\alpha} - \frac{1}{4}  g_{\mu\nu} \mathbf{F}^{~\alpha\beta} \mathbf{F}_{\alpha\beta}
\right)\,,\\
 \mathbf{T}_{\mu\nu}^{\Phi} &=&
  \frac{1}{16 \pi  G } \left(
\partial_\mu \mathbf{\Phi} \partial_\nu \mathbf{\Phi} - \frac{1}{2} |\partial \mathbf{\Phi}|^2 g_{\mu\nu}
\right)\,.
\end{eqnarray}
So the equation of the motion of the metric from  \eqref{metricGTall} indeed converts to:
\begin{equation}
G_{\mu\nu} = 8 \pi G T^{M}_{\mu\nu}\,,
\end{equation}
which is the Einstein-Hilbert equation for the metric. This proves that  imposing \eqref{ConsCoupling} in the EH regime of  \eqref{Ldefition} at the classical level reproduces the Einstein-Hilbert gravity, where the gauge fields and scalars affect neither the space-time geometry nor the orbits of particles/probes.  

Eq. \eqref{ConsCoupling} implies that one of the gauge fields abs one of the scalars have the kenitic energy term of ghosts. The theory, however, is equivalent to the Einstein-Hilbert theory.  To see this define:
\begin{eqnarray}
A^\pm = \frac{1}{2} (A_\mu^{1}\pm A^{2})\,,\\
\phi^\pm = \frac{1}{2} (\phi^{1}\pm \phi^{2})\,, 
\end{eqnarray}
Rewrite the action in term of $A^\pm_\mu$ and $\phi^\pm$:
\begin{eqnarray}
S[A_1]+S[A_2] &=&  \frac{1}{8 \pi \kappa_1 G}\int d^4 x  \sqrt{-\det g} \, A^-_\nu \partial_\mu F^{+\mu\nu} \,,\\
S[\phi_1]+S[\phi_2] &=&  \frac{1}{8 \pi \tilde{\kappa}_1 G}\int d^4 x  \sqrt{-\det g} \, \phi^-\Box \phi^+\,,
\end{eqnarray}
The exact Feynman path integral (up to a normalisation factor), then, is simplified to:
\begin{eqnarray}\label{AAZ}
Z&=& \int Dg DA^+ DA^- D\phi^+ D\phi^-e^{-\frac{iS}{\bar{h}}}\\
&=& \int Dg  DA^+ D\phi^+  \boldsymbol{\delta}(\partial^\mu F^+_{\mu\nu}) \boldsymbol{\delta}(\Box \phi^+) e^{-\frac{i}{\hbar} (S[g]+S_M) }\,,
\end{eqnarray}
 where the integrations over $A^-$ and $\phi^-$ are performed. The simplified path integral gives the exact equations of $A^+$ and $\phi^+$ at the level of quantum field theory:
 \begin{eqnarray}
 \partial^\mu F^+_{\mu\nu}&=&0\,,\\
 \Box \phi^+ &=& 0 \,,
 \end{eqnarray}
 which are solved by
 \begin{eqnarray}
 A^+_\mu &\equiv& 0, \\
 \phi^+ &\equiv & 0\,,
 \end{eqnarray}
 inserting which into \eqref{AAZ} reproduces the exact Feynman path integral of the Einstein-Hilbert action. This proves the EH limit of the theory coincides to the Einstein-Hilbert theory at the level of quantum field theory too.  

\section{Static Solutions}
\label{staticsolutions}
Let a  static mass distribution be considered in an asymptotically flat space-time.
The metric in the not-strong regime of the Einstein-Hilbert part of the action reads:
\begin{equation}
\label{metricSC}
ds^2= - (1+ 2 \phi_N ) dt^2 + (1- 2 \phi_N ) d\vec{x}^2\,,
\end{equation}
where $\phi_N$ holds
\begin{equation}
\nabla^2 \phi_N= 4 \pi G \rho \,,
\end{equation}
and $\rho$ is the density, and $2 \phi_N\ll 1$.
The gauge fields and scalars can be represented by:
\begin{eqnarray}
A_{1\mu}  &=& ( A_1(x),0,0,0) ,\\
A_{2\mu}  &=& ( A_2(x),0,0,0) ,\\
\phi_1 &=& \phi_1(x)\,,\\
\phi_2 &=& \phi_2(x)\,.
\end{eqnarray}
The four  vector velocity field of the matter distribution reads:
\begin{equation}
u^\mu = (u(x),0,0,0)\, ,
\end{equation}
Notice that \eqref{umuconst} then implies
\begin{equation}
u(x) =  \frac{e^{-\bar{\phi}(x)}}{1+\phi_N}\,.
\end{equation}
where $\bar{\phi}(x) = \phi_{1}(x)+\phi_{2}(x)$.  Utilizing this metric in \eqref{ScalarsEquations}, and using the identity of
\begin{equation}
 \Box \phi = \frac{1}{\sqrt{-\det g}}\partial_\mu (\sqrt{-\det g} g^{\mu\nu} \partial_\nu \phi)\,,
 \end{equation}
 and recalling that $\phi_N\ll 1$, returns
  \begin{eqnarray}
  \label{phi1sol}
 \phi_1 &=&4 \tilde{k}_1  \phi_N \,,\\
 \phi_2 &=& 4 \tilde{k}_2 \phi_N  \,,
 \end{eqnarray}
The equations of the gauge fields in the Einstein-Hilbert regime gives:
\begin{eqnarray}
\label{A1sol}
 A_1 &=&4 {k}_1  \phi_N \,,\\
 A_2 &=& 4 {k}_2 \phi_N  \,,
\end{eqnarray}
where the boundary conditions are imposed such that the fields and their first derivatives vanish for vanishing total mass.

 Although the net contribution of the gauge fields and scalars  to the energy momentum tensor does not vanish in the MOND regime, it is negligible assuming. The metric in the MOND regime, therefore, coincides to  \eqref{metricSC} at the leading order approximation and can be approximated to the flat Minkowski metric. This simplifies the equation of motion in the MOND regime  -\eqref{gaugeequationsb} and \eqref{ScalarsEquationsb} for static solutions to:
\begin{eqnarray}
\nabla^i (\frac{l |\nabla \phi_2|}{\sqrt{2}} \nabla_i \phi_2) &=& 16 \pi \tilde{\kappa}_2 G \rho\,,\\
\nabla^i (\frac{l |\nabla A_2|}{\sqrt{2} } \nabla_i A_2) &=& 16 \pi {\kappa}_2 G \rho\,,
\end{eqnarray}
Recall the AQUAL equation in its MOND regime \cite{AQUAL}:
\begin{equation}
\nabla_i \left( \frac{|\nabla \phi_{m}|}{a_\star} \nabla^i \phi_{m} \right)= 4 \pi G \rho\,.
\end{equation}
So we can write:
\begin{subequations}
\label{A2phi2}
\begin{eqnarray}
\phi_2(r) &=& 2\, \text{sign}({\tilde{\kappa}_2}) \sqrt{|\tilde{\kappa}_2|} \, \phi_m\,, \\
A_2(r) &=& 2\, \text{sign}({\kappa}_2) \sqrt{|\kappa_2|}\, \phi_m \,,
\end{eqnarray}
\end{subequations}
where:
\begin{eqnarray}
\label{defa0}
a_\star &\equiv& \frac{c^2}{\sqrt{2} l}\,.
\end{eqnarray}

\subsection{Orbit of particles}
\label{OrbitOfParticles}
Eq. \eqref{Randers.3} that governs orbits of massive particle for static mass distribution simplifies to:
\begin{eqnarray}
\label{orbit1}
S[x(\tau)] & =&  \int d\tau L \\
L&=&
+ \frac{m e^{2 \bar{\phi}} }{2}  (-(1+ 2\phi_N) \dot{t}^2 + (1-2\phi_N) |\dot{\vec{x}}|^2)
-m \bar{A}_0 \dot{t}\,.
\end{eqnarray}
For a slow moving massive particle, one can choose
\begin{eqnarray}
\tau &=& t\,,
\label{t=tau}
\\
|\dot{\vec{x}}| &\ll & 1\,,
\end{eqnarray}
that simplifies \eqref{orbit1} to
\begin{equation}
S = \int dt \,(\frac{1}{2} m |\dot{\vec{x}}|^2 - m (\bar{A}_0+ \bar{\phi}+ \phi_N))+O(\bar{\phi}^2,  \phi_N \bar{\phi}, \phi_N^2)\,.
\end{equation}
So the effective weak gravitational potential for a slow moving probe follows
\begin{equation}\label{phislow}
\phi_{eff}^{slow}\,=\,\bar{A}_0+ \bar{\phi}+ \phi_N\,.
\end{equation}
Now let a fast moving particle be considered. We can choose the following coordinate
\begin{subequations}
\label{DiracCordinates}
\begin{eqnarray}
t & =& t \,,\\
x_3 & =& t + \tilde{x}_3\,,\\
x_1 & = & \tilde{x}_1\,,\\
x_2 & = & \tilde{x}_2\,,
\end{eqnarray}
\end{subequations}
with the world-line parametrization of \eqref{t=tau} such that
\begin{eqnarray}
|\dot{\tilde{x}}_1|,|\dot{\tilde{x}}_2|,|\dot{\tilde{x}}_3|&\ll & 1\,.
\end{eqnarray}
Inserting \eqref{DiracCordinates} into \eqref{Randers.3} gives
\begin{equation}
S = \int dt \,(\frac{1}{2} m |\dot{\vec{\tilde{x}}}|^2 - m (\bar{A}_0+ 2\bar{\phi}))+O(\bar{\phi}^2, \phi_N \bar{\phi}, \phi_N^2)\,.
\end{equation}
which means that the effective weak gravitational potential for a fast moving probe including a light signal reads
\begin{equation}\label{phifast}
\phi_{eff}^{fast}\,=\,\bar{A}_0+ 2 \phi_N\,.
\end{equation}
Notice that as expected, $\bar{\phi}$ does not affect  the light's trajectory.
The Newtonian regime due to $\bar{A}_0=\bar{\phi}=0$ holds
\begin{equation}
\frac{ \phi_{eff}^{fast}}{ \phi_{eff}^{slow}}\,=\,\gamma+1 = \,2\,,
\end{equation}
which means that the post Newtonian parameter of $\gamma=1$. Since the MOND regime also holds $\gamma=1$ \cite{Amendola:2013qna} then we conclude from \eqref{phislow} and \eqref{phifast} that the MOND regime should hold
\begin{equation}
\label{barphiA2}
\bar{\phi}=- \frac{1}{2} \bar{A}_0\,,
\end{equation}
which means that the scalars decrease the value of effective gravitational potential to $50\%$.
We notice that  \eqref{barphiA2} can not be hold in the interpolating regime: the regime that connects the Newtonian regime to the MONDian regime.  We discuss the MOND regime in the next section. Einstein-Hilbert regime holds \eqref{barphiA2}. In the MOND regime, after approximating $\bar{A}_0$ and $\bar{\phi}$  respectively to $\bar{A}_2$ and $\bar{\phi}_2$,   \eqref{barphiA2}  demands
\begin{equation}
\tilde{\kappa}_2 = -\frac{1}{4} \kappa_2 \,,
\end{equation}
which beside \eqref{ConsCoupling} allows to express all the coupling constants in term of a constant:
\begin{eqnarray}
\kappa_1 &=& -\kappa\,,\\
\kappa_2 &=& +\kappa\,,\\
\tilde{\kappa}_1 &=&+\frac{1}{4} \kappa\,,\\
\tilde{\kappa}_2 &=&-\frac{1}{4} \kappa\,.
\end{eqnarray}
Far from the mass distribution  holds
\begin{subequations}
\label{phiNphim}
\begin{eqnarray}
\partial_r \phi_N &=&  \frac{GM}{r^2} \hat{r} +O(\frac{1}{r^3})\,,\\
\partial_r \phi_m &=&  \frac{\sqrt{GM a_\star}}{r} \hat{r}+ O(\frac{1}{r^2}) \,,
\end{eqnarray}
\end{subequations}
where $M$ represents the total mass.  Inserting \eqref{phiNphim} into \eqref{phi1sol}, \eqref{A1sol} and \eqref{A2phi2}, and using the results in \eqref{phislow}  gives
\begin{eqnarray}
\label{phislowasmp}
\partial_r \phi^{slow}_{eff}=  \,  \text{sign}(k) \frac{\sqrt{GM |\kappa| a_\star} }{r} + O(\frac{1}{r^2})\,.
\end{eqnarray}
We notice that requiring the gravitational attraction at large $r$ demands that
\begin{equation}
\kappa > 0 \,.
\end{equation}
In order to reproduce the flat rotational velocity curves of spiral galaxies, the effective potential should coincide to \cite{AQUAL}:
\begin{subequations}
\label{a0Obserations}
\begin{eqnarray}
 \partial_r \phi^{slow}_{eff}&=& \frac{\sqrt{GM a_{o}}}{r}\hat{r}+ O(\frac{1}{r^2})\,,\\
 a_{o} &=& (1\pm 0.2) \times 10^{-10}\frac{m}{s^2}\,.
\end{eqnarray}
\end{subequations}
The consistency between \eqref{defa0}, \eqref{phislowasmp} and  \eqref{a0Obserations} then demand:
\begin{equation}\label{loverk}
\frac{l}{\kappa}= \frac{ c^2}{\sqrt{2}a_o}\,.
\end{equation}

\section{Weak Gravitational Lensing}
\label{WeakGravitationalLensing}
\begin{figure}[tbp]
        \centering
               \begin{subfigure}[b]{\textwidth}
                \centering
                \includegraphics[width=0.45\textwidth]{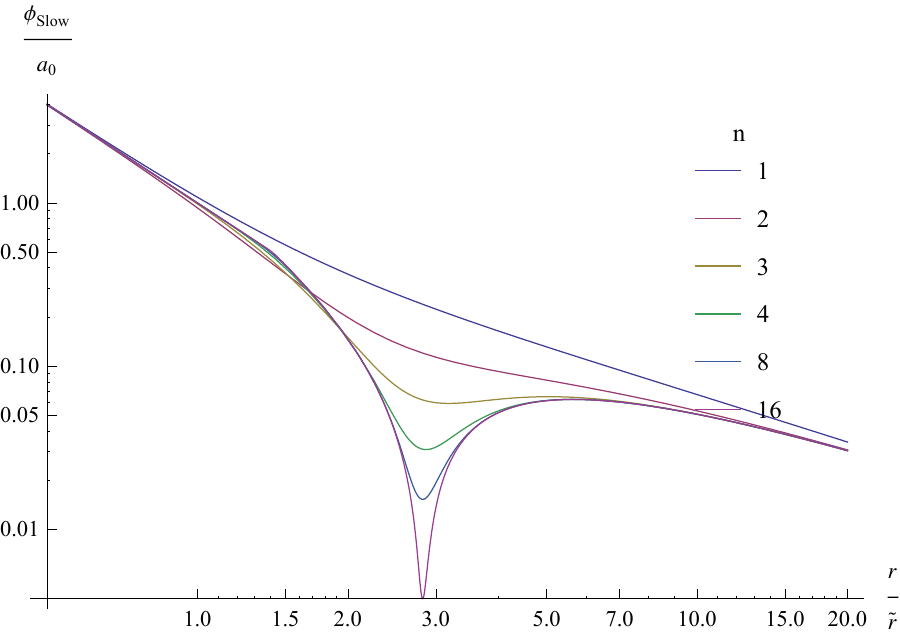}
                \includegraphics[width=0.45 \textwidth]{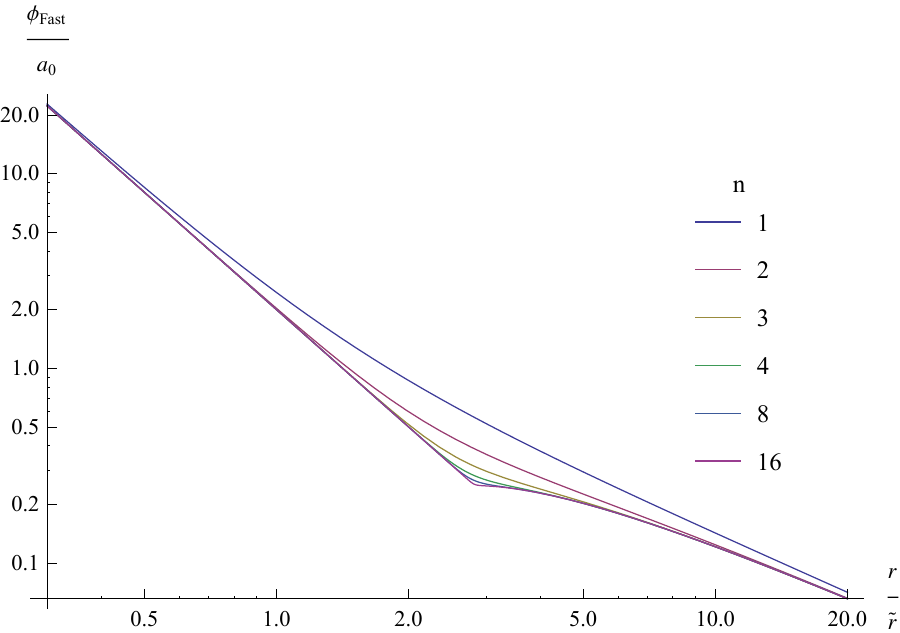}
                \caption{$\kappa=1$}
         \end{subfigure}
        \begin{subfigure}[b]{\textwidth}
                \centering
                \includegraphics[width=0.45\textwidth]{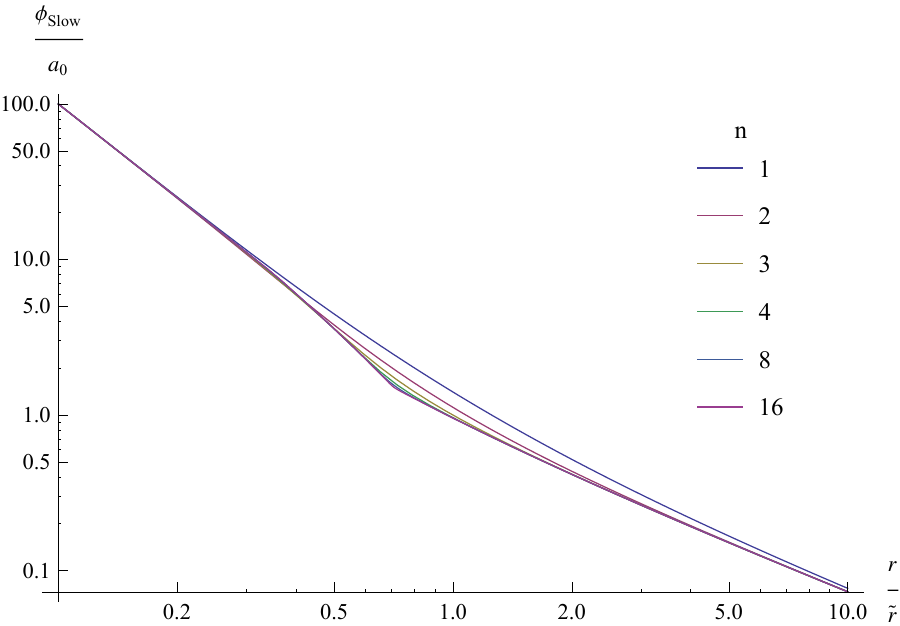}
                 \includegraphics[width=0.45\textwidth]{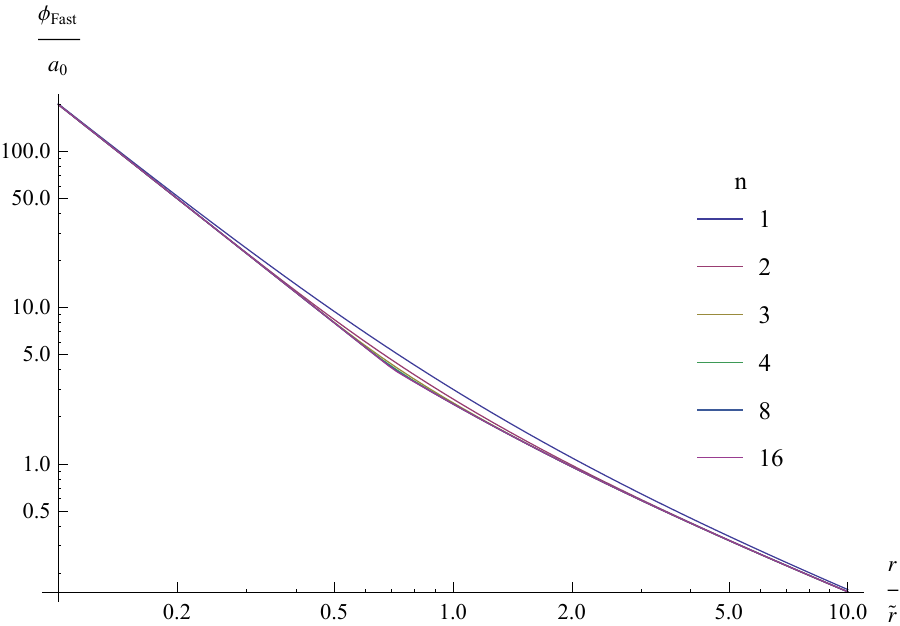}
                \caption{$\kappa=\frac{1}{4}$}
        \end{subfigure}
        \begin{subfigure}[b]{\textwidth}
                \centering
                \includegraphics[width=0.45\textwidth]{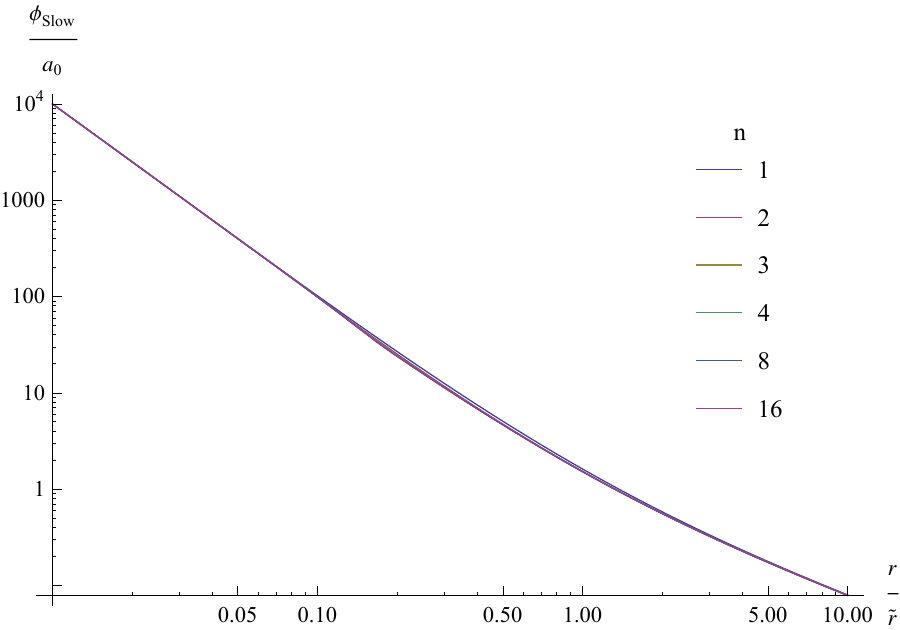}
                \includegraphics[width=0.45\textwidth]{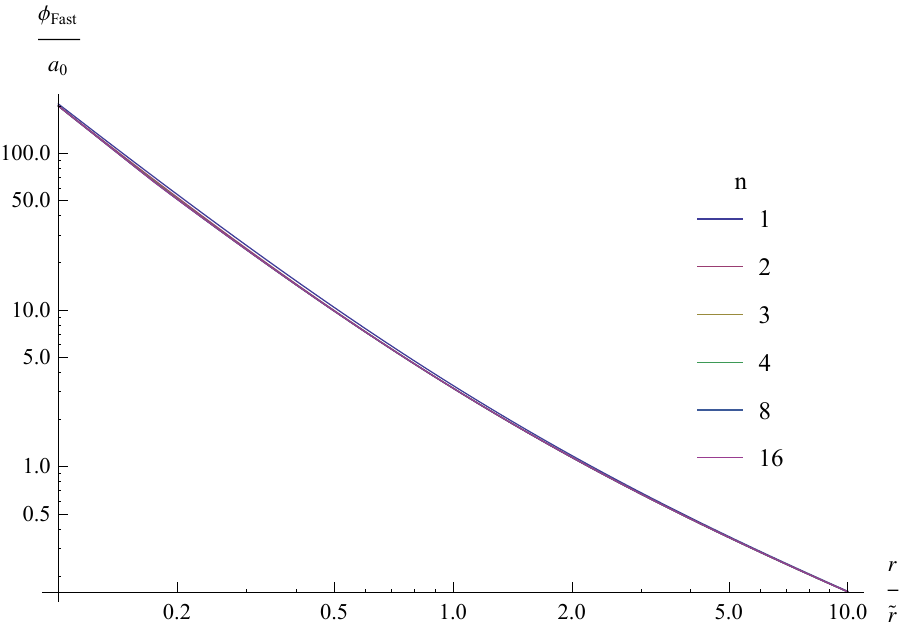}
                \caption{$\kappa=\frac{1}{16}$}
         \end{subfigure}
            \caption{The  effective gravitational acceleration  felt by a slow (left) and fast (right) moving probe for various values of $\kappa$ and different choices of the interpolating function.}
        \label{fig:2}
\end{figure}

Since $l$ is  \eqref{loverk} sufficiently large,  we can approximate  \eqref{metricSC}  to the  Minkowsky metric and simplify \eqref{ScalarsEquationsb} and \eqref{gaugeequationsb} around a spherical static mass distribution to:
\begin{subequations}
\label{gammaA2phi2}
\begin{eqnarray}
{\cal L}'(\frac{l^2}{2 c^4} |\partial_r A_2|^2)\partial_r A_2 &=& 4 \kappa \frac{G M}{r^2}\,,\\
{\cal L}'(\frac{l^2}{2 c^4} |\partial_r \phi_2|^2)\partial_r \phi_2 &=& -\kappa \frac{G M}{r^2}\,,
\end{eqnarray}
\end{subequations}
where $M$ is the total mass. To find the exact value of the fields we consider the following family of the interpolating functions:
\begin{eqnarray}
\label{mun}
{\cal L}'(x^2) &=& \mu_n(x)= \frac{x}{(1+x^n)^\frac{1}{n}}\,,
\end{eqnarray}
where $x>0$. We can write
\begin{eqnarray}
\label{ln}
{\cal L}_n(x) &=& \frac{2}{3} x^\frac{3}{2} ~{}_2F_1(\frac{1}{n},\frac{3}{n},1+\frac{3}{n},-x^{\frac{n}{2}})+c_1\,,
\end{eqnarray}
where ${}_2F_1$ is the Gauss's hypergeometric function and $c_1$ is the constant of integration. Notice that
\begin{equation}
\lim_{x\to 0} {\cal L}_n(x) = c_1\,,
\end{equation}
 and $c_1$ is tantamount to adding a cosmological constant. For the time being we set
 \begin{equation}
 c_1 \,=\,0 \,.
 \end{equation}
 The first few  ${\cal L}_n(x)$ follows
\begin{eqnarray}
{\cal L}_1(x) &=& - 2\sqrt{x}+ x + 2 \ln(1+\sqrt{x})\,,\\
{\cal L}_\frac{3}{2}(x) & = & \sqrt[3]{x^{3/4}+1} (x^{3/4} -3 )+3\,,\\
{\cal L}_2(x) &=&\sqrt{x(1+x)}-\text{ArcSinh}(\sqrt{x})\,,\\
{\cal L}_3(x) &=&  -1+ (1+ x^{\frac{3}{2}})^{\frac{2}{3}}\,,
\end{eqnarray}
The literature knows the cases of $n=1$ and $n=2$  respectively  as the simple and the standard choices. Define new variables
\begin{subequations}
\label{auxA2phi2Static}
\begin{eqnarray}
x_a &\equiv& \frac{l}{\sqrt{2} c^2} \partial_r A_2\,,\\
x_\phi &\equiv& -\frac{l}{\sqrt{2} c^2} \partial_r \phi_2 \,,
\end{eqnarray}
\end{subequations}
and insert them into \eqref{gammaA2phi2}:
\begin{eqnarray}
\frac{x_\phi^2}{(1+x_\phi^n)^{\frac{1}{n}}} &=& 2\kappa^2  \frac{\tilde{r}^2}{r^2}\,, \\
\frac{x_a^2}{(1+x_a^n)^{\frac{1}{n}}} &=& 8 \kappa^2  \frac{\tilde{r}^2}{r^2}\,,
\end{eqnarray}
where \eqref{loverk} is ultilized and
\begin{eqnarray}
\tilde{r} \equiv \frac{1}{2}\sqrt{\frac{GM}{a_o}}\,.
\end{eqnarray}
\eqref{auxA2phi2Static} is  solved by:
\begin{eqnarray}
\label{xphi}
x_\phi &=& 2^{^{\frac{n-2}{2n}}}\, \frac{\kappa \tilde{r}}{r}
\left(
 2^{{\frac{n}{2}}} (\frac{\kappa \tilde{r}}{r})^n +  \sqrt{4+  2^{n}(\frac{\kappa \tilde{r}}{r})^{2n}}
\right)^{\frac{1}{n}}\,,\\
\label{xa}
x_a &=&2^{^{\frac{3}{2}-\frac{1}{n}}}  \frac{ \kappa \tilde{r}}{r}
\left(
 8^{{\frac{n}{2}}} (\frac{\kappa \tilde{r}}{r})^n + \sqrt{4+ 8^{n} ( \frac{\kappa \tilde{r}}{r})^{2n}}
\right)^{\frac{1}{n}} \,,
\end{eqnarray}
which utilizing \eqref{auxA2phi2Static} gives
\begin{eqnarray}
\label{phi2xphi}
\phi_2' &=&-\frac{a_0}{2 \kappa} x_\phi,\\
\label{A2xa}
 A_2' &=& \frac{a_o}{2 \kappa} x_a.
\end{eqnarray}
Eq. \eqref{phi1sol} and \eqref{A1sol} give:
\begin{eqnarray}
\phi_1' &=&  \kappa a_o\frac{\tilde{r}^2}{r^2}\\
A_1' &=& - 4 \kappa a_o\frac{\tilde{r}^2}{r^2}
\end{eqnarray}
Now let $\tilde{\gamma}$ be defined by
\begin{equation}
\tilde{\gamma} \equiv \frac{\partial_r \phi_{eff}^{fast}}{\partial_r \phi_{eff}^{slow}}  = \frac{\bar{A}_0'+ 2 \phi_N'}{\bar{A}'_0 + \bar{\phi}'+\phi_N'}\,,
\end{equation}
in the l. h. s. of which \eqref{phislow} and \eqref{phifast} are utilized and the prime superscript  represents differentiation  with respect to $r$.  The Einstein-Hilbert gravity requires that $\tilde{\gamma}=1$, and in or study it is easier to work with $\tilde{\gamma}$ rather than $\gamma$. The exact form of $\tilde{\gamma}$ for any choice of $\mu(n)$ in \eqref{mun} follows:
\begin{figure}[tbp]
        \centering
               \begin{subfigure}[b]{0.45\textwidth}
                \centering
                \includegraphics[width=\textwidth]{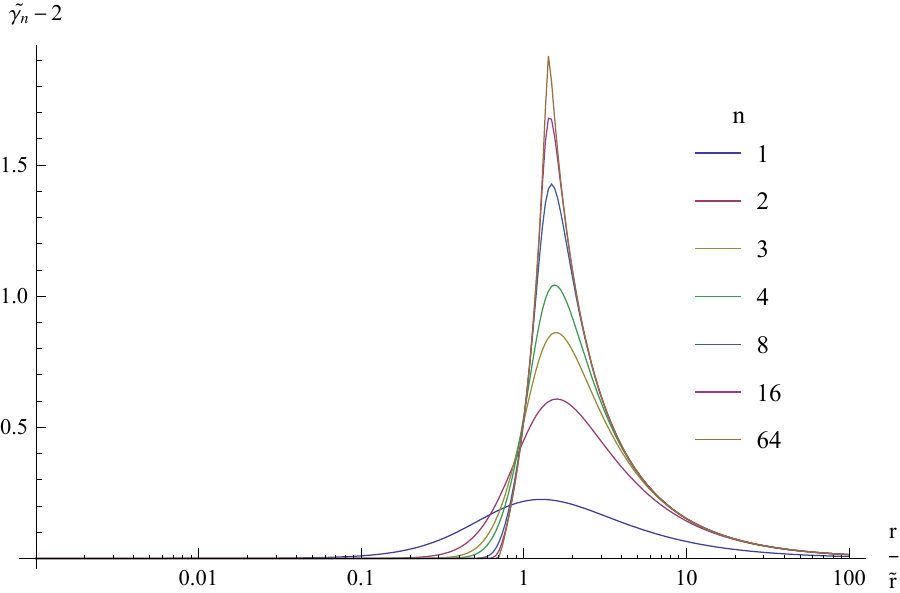}
                \caption{$\kappa=\frac{1}{2}$ }
        \end{subfigure}%
        ~ 
        \begin{subfigure}[b]{0.45\textwidth}
                \centering
                \includegraphics[width=\textwidth]{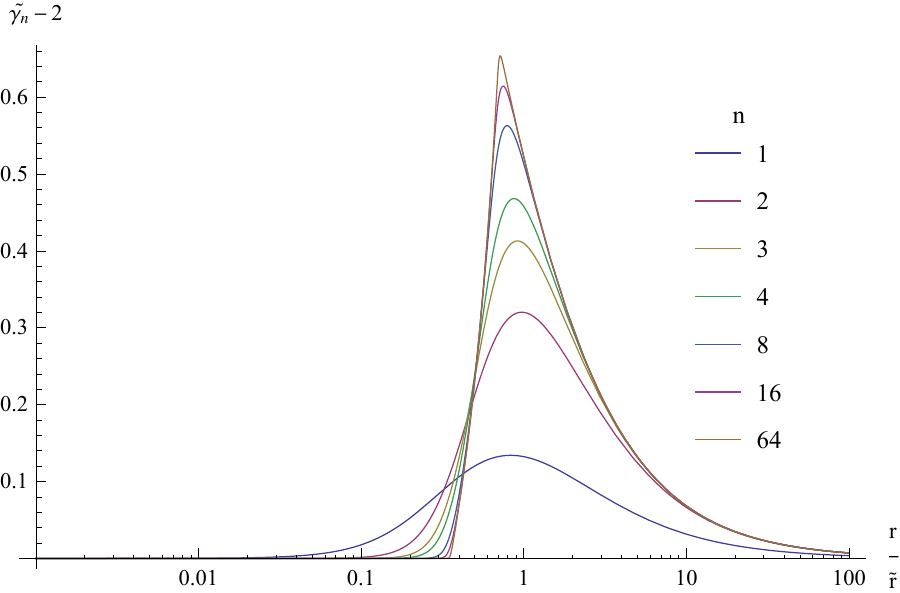}
                \caption{$\kappa=\frac{1}{4}$}
        \end{subfigure}
        \begin{subfigure}[b]{0.45\textwidth}
                \centering
                \includegraphics[width=\textwidth]{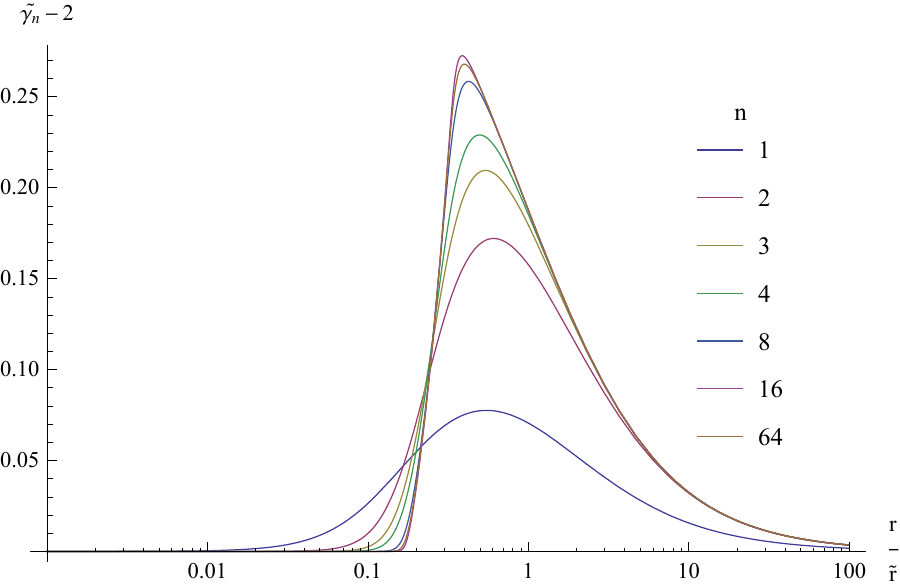}
                \caption{$\kappa=\frac{1}{8}$ }
        \end{subfigure}%
        ~ 
        \begin{subfigure}[b]{0.45\textwidth}
                \centering
                \includegraphics[width=\textwidth]{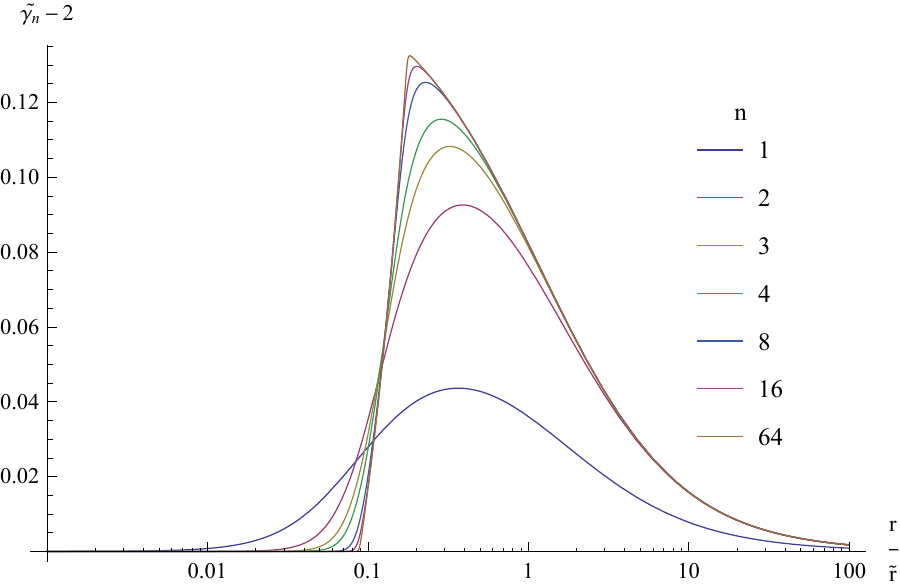}
                \caption{$\kappa=\frac{1}{16}$}
        \end{subfigure}
        \begin{subfigure}[b]{0.45\textwidth}
                \centering
                \includegraphics[width=\textwidth]{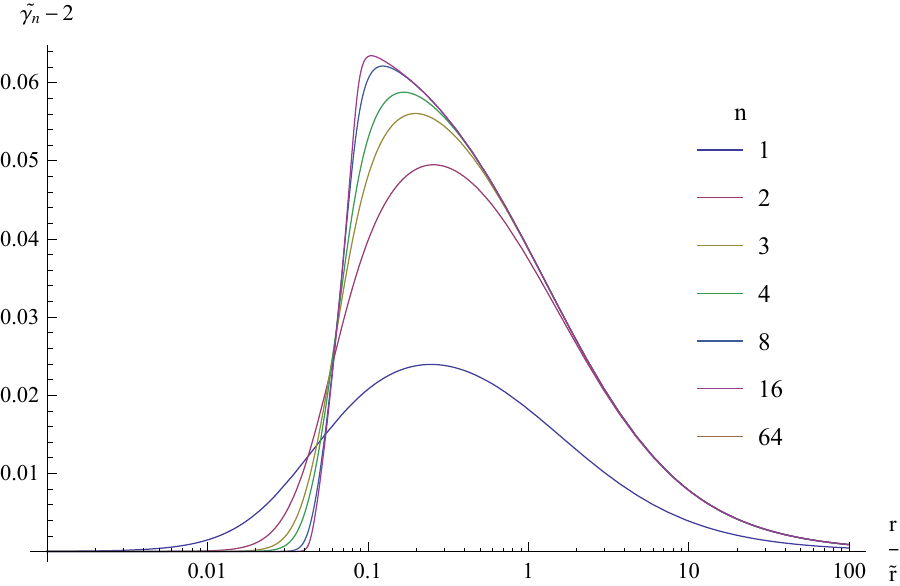}
                \caption{$\kappa=\frac{1}{32}$}
        \end{subfigure}%
        ~ 
        \begin{subfigure}[b]{0.45\textwidth}
                \centering
                \includegraphics[width=\textwidth]{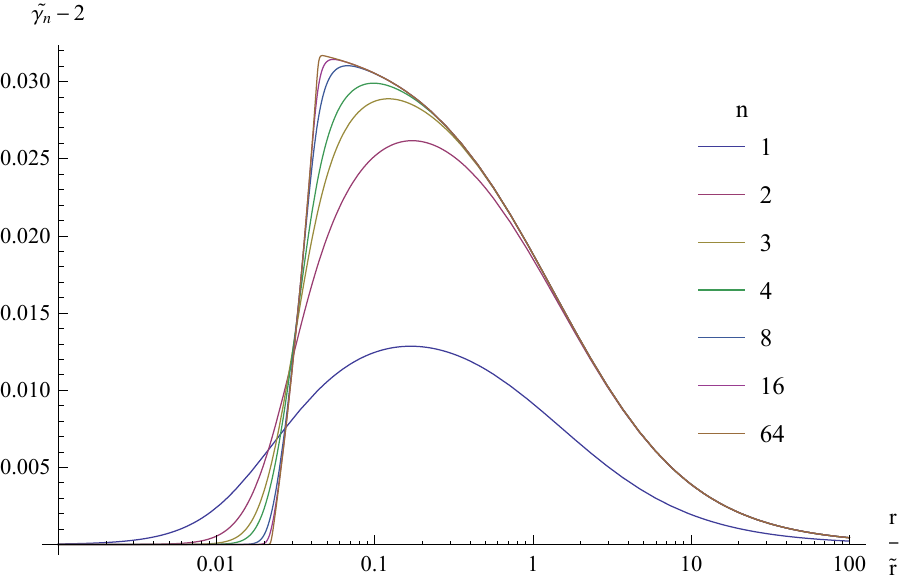}
                \caption{$\kappa=\frac{1}{64}$}
        \end{subfigure}
            \caption{The  $\tilde{\gamma}$ for various values of $\kappa$ and different choices of the interpolating function.}
        \label{fig:1}
\end{figure}
\begin{eqnarray}
\frac{1}{\tilde{\gamma}_n}= 1+\frac{-2^{1+\frac{1}{n}} (-1+\kappa)+\frac{\sqrt{2}r}{\tilde{r}} \left(\sqrt{4+2^n \left(\frac{\kappa\tilde{r}}{r}\right)^{2 n}}+2^{n/2} \left(\frac{\kappa\tilde{r}}{r}\right)^n\right)^{\frac{1}{n}}
}{2^{2+\frac{1}{n}} (-1+2 \kappa)-\frac{2 \sqrt{2}r}{\tilde{r}} \left(\sqrt{4+8^n \left(\frac{\kappa\tilde{r}}{r}\right)^{2 n}}+2^{3 n/2} \left(\frac{\kappa\tilde{r}}{r}\right)^n\right)^{\frac{1}{n}}
}\,.
\end{eqnarray}
Fig. \ref{fig:1} depicts $\tilde{\gamma}_n$ for various values of $n$ and $\kappa$. As depicted in the fig. \ref{fig:2} , the sharp increase in $\tilde{\gamma}$ is due to the sharp decrease of $\phi_{eff}^\text{slow}$.
 \begin{figure}[tbp]
        \centering
                \includegraphics[width=0.7 \textwidth]{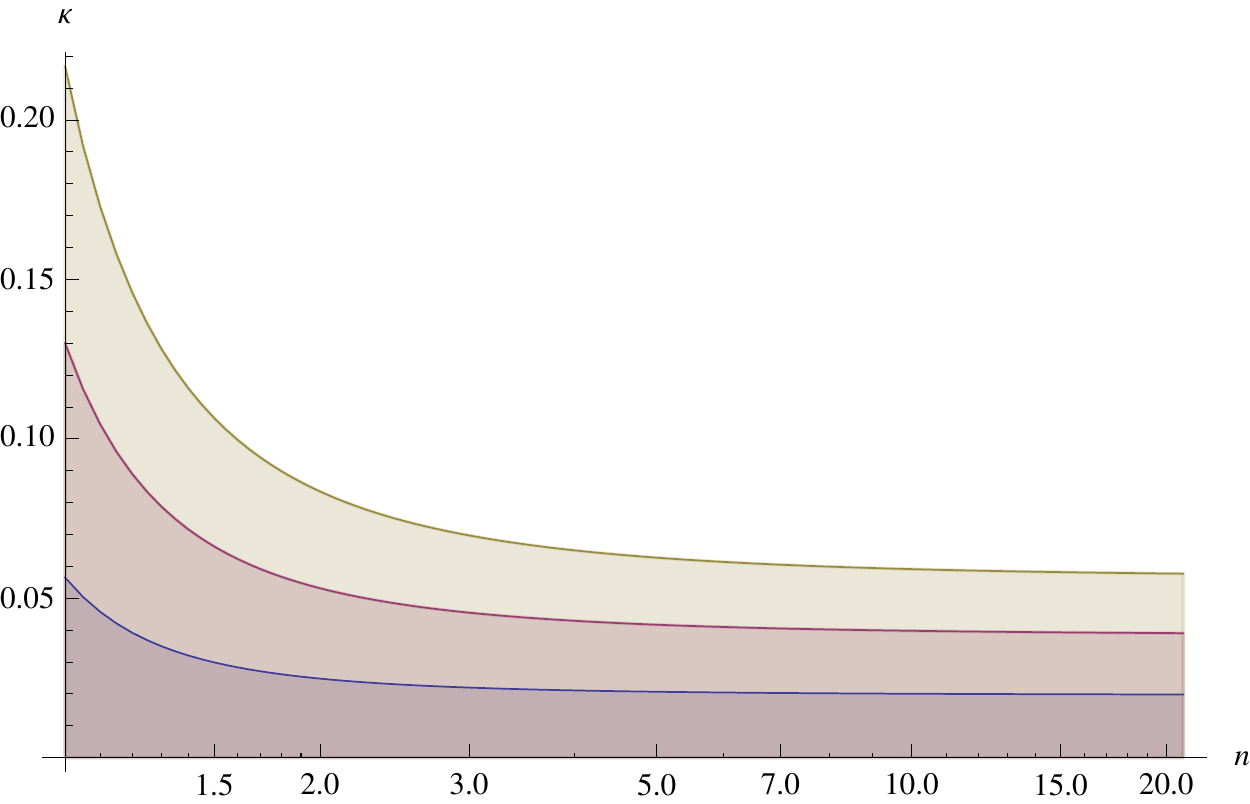}
            \caption{The colored regimes shows the values of $\kappa$ and $n$ that are consistent with the weak gravitational lensing at one,  two and three sigma confidence levels.}
        \label{knplotWGL}
\end{figure}
We see that requiring $\tilde{\gamma}$ to be close to $2$ sets non-trivial constraints on $\kappa$ and $n$.  Ref. \cite{Amendola:2013qna} reports that
\begin{eqnarray}
\frac{|\gamma-1|}{2} \le 0.02\,,
\end{eqnarray}
at one sigma level. This means that $\gamma$ is constant at the leading order approximation. Therefore,  it equivalently holds
\begin{eqnarray}\label{gammatildeconstraint}
\frac{|\tilde{\gamma}_n-2|}{2} \le 0.02\,.
\end{eqnarray}
We notice that the maximum value of $\gamma$ holds $\gamma'_n(r)=0$. So
\begin{eqnarray}
\tilde{\gamma}_n(r) &=& 2+ \text{CL}\times 0.04 \,,\\
\tilde{\gamma}'_n(r)&=&0
\end{eqnarray}
 can numerically be solved in order to find  the maximum value of $\kappa$ that holds \eqref{gammatildeconstraint} everywhere with the confidence level of CL.
 Fig. \ref{knplotWGL} depicts the values of $\kappa$ and $n$ holding   \eqref{gammatildeconstraint} everywhere at one, two and three sigma confidence levels.  For these allowed values, no sharp decrease exists in the effective potentials.

\begin{figure}[tbp]
        \centering
                \includegraphics[width=0.7 \textwidth]{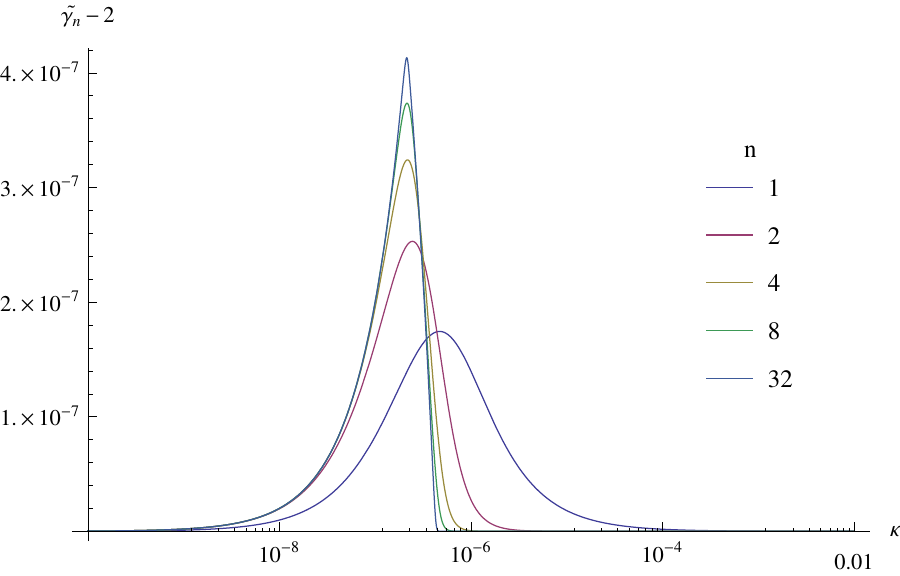}
            \caption{$\tilde{\gamma}_n-2$ at the surface of the Sun for various values of $n$ and $\kappa$.}
        \label{fig:SunGamma}
\end{figure}

The $\gamma$ parameters is measured in strong gravity too. One of the best bound on its strong gravity value reads \cite{StrongG1,StrongG2}:
\begin{equation}\label{GammaStongObservation}
\gamma -1  = (0.8\pm 1.2) \times 10^{-4}\,,
\end{equation}
Fig. \ref{fig:SunGamma} depicts the value of $\tilde{\gamma}_n$ at the surface of the Sun for various values of $\kappa$ and $n$. We observe that they are in agreement with \eqref{GammaStongObservation}.

\section{Lunar System Constraints}
\label{LunarSystemConstraints}
The effective $GM$ is defined by
\begin{equation}
GM^{eff}(r)= r^2 \partial_r \phi_{eff}^{slow}\,.
\end{equation}
 The accurate value of the mass ratio of the Sun/(Earth+Moon) from the Lunar Laser Ranging combined with the Solar GM and the lunar GM from lunar orbiting spacecrafts \cite{LLR} yields the effective   gravitational mass of the Earth in an Earth-centered reference frame with the precision of one part in $10^8$ :
\begin{equation}\label{GMLLR}
GM^{LLR}_{\oplus}(r_{_{LD}}) = 398600.443 \pm 0.004
\frac{km^3}{s^2}
\end{equation}
where $r_{_{LD}}$ represents the Lunar distance: the average distance  between the Earth and Moon:
\begin{equation}
r_{_{LD}}= 384,400 km\,,
\end{equation}
The  Earth GM has also been measured by various artificial Earth satellites, including the accurate tracking of the LAGEOS
satellites orbiting the Earth in nearly circular orbits with
semimajor axes about twice the radius of the Earth \cite{LAGEOS}:
\begin{equation}\label{GMLAGEOS}
GM^{LAGEOS}_{\oplus}(2 r_{\oplus})= 398600.4419\pm 0.0002    \frac{km^3}{s^2}
\end{equation}
where $r_{\oplus}$ stands for the radius of the Earth:
\begin{equation}
r_{\oplus} = 6,371 km\,.
\end{equation}
The consistency with the Lunar system, thus, requires
\begin{equation}
\label{LunarConstraint}
|\frac{GM^{eff}_\oplus(r_{_{LD}}) - GM^{eff}_\oplus(2r_\oplus)} {GM_\oplus} |< 10^{-8}\,.
\end{equation}
We notice that the distance between the Moon and the earth are measured by sending light signals. Deviation from the Einstein-Hilbert geometry affects the time that a signal travel from the Earth to the moon and back. $\bar{\phi}$ does not contribute to null geodesics. Our theory adds $\Delta T$ to the time of travel for the signal:
\begin{equation}
\label{DeltaTLunar}
\Delta T = \frac{2}{c^2} \int_{r_\oplus}^{r_{_{LD}}}d\tau \bar{A} \dot{t} \approx \frac{2}{c^3} \int_{r_\oplus}^{r_{LD}} dr \bar{A} = \frac{2}{c^3} \int_{r_\oplus}^{r_{_{LD}}} dr (\phi_{eff}^{fast}-2\phi_N) \,,
\end{equation}
wherein we have approximated $dt =\frac{1}{c} dr$ since the space-time geometry around the Earth is almost flat. We can taylor expands the fields in the Newtonian regime of the Lunar system,, in the regime that holds $(\frac{k \tilde{r}}{r})^{2}\gg 1 $\footnote{$n=1$ does not resolve the Pioneer anomaly.}:
\begin{eqnarray}
\label{phiLunarExpansion}
\partial_r \phi_{eff}^{slow} &=& \frac{G M}{r^2}
\left(
1- \frac{2^{2n}-4}{2^{3n}}\frac{\kappa}{n} (\frac{\kappa\tilde{r}}{r})^{-2n}
+\cdots
\right) \,,\\
\label{DPhiFastLeading}
\partial_r \phi_{eff}^{fast} &=& \frac{2G M}{r^2}
\left(
1
+ 2^{1-3n}\frac{k}{n} (\frac{\kappa\tilde{r}}{r})^{-2n}
+\cdots
\right) \,,
\end{eqnarray}
 In order to have a consistent Taylor series for the Lunar system, we demand that
\begin{subequations}
\label{k1lunar}
\begin{eqnarray}
\label{lunarExpansionAllowed}
(\frac{\kappa \tilde{r}_\oplus}{\sqrt{8}r_{_{LD}}})^{-2n} &<& {10^{-p}}\,, \to \\
 \kappa_1(n,p) &\equiv & 8\sqrt{2} \times 10^{\frac{p}{2n}-4}\,,\\
 \kappa_1(n,p) &<& \kappa\,,
\end{eqnarray}
\end{subequations}
where $p\ge0$ and the larger $p$ is the better the Taylor series is.  Eq. \eqref{DPhiFastLeading} leads to
\begin{equation}
c \Delta T = \frac{2^{1-3n} \kappa^{1-2n}}{n^2(2n-1)} \left((\frac{r_{LD}}{\tilde{r}_\oplus})^{2n}-(\frac{2r_\oplus}{\tilde{r}_\oplus})^{2n}\right) \frac{G M_\oplus}{c^2}\,,
\end{equation}
wherein $\phi_{eff}^{fast}- 2\phi_N$ is assumed to vanish at $r=0$. Notice that
\begin{eqnarray}
\frac{r_{LD}}{\tilde{r}_\oplus} &=&  4\times 10^{-4}\,,\\
\frac{2r_\oplus}{\tilde{r}_\oplus} &=&  1.2 \times 10^{-5}\,,\\
\frac{G M_\oplus}{c^2} &=& \frac{9}{2} \times 10^{-3} m\,.
\end{eqnarray}
The LLR measurements identify the center of the earth with the precision of few centimeters.  We require
\begin{equation}
\label{cDeltaTLunar}
c \Delta T < 1 cm\,,
\end{equation}
which induces the following lower bound on $\kappa$:
\begin{subequations}
\label{k2lunar}
\begin{eqnarray}
\kappa_2(n)&\equiv&\left(\frac{9\times 2^{n}}{n^2(2n-1)}\right)^{\frac{1}{2n-1}} 10^{-4-\frac{3}{2n-1}} \,,\\
\kappa_2(n)&<& \kappa\,.
\end{eqnarray}
\end{subequations}
\begin{figure}[tbp]
               \centering
               \begin{subfigure}[b]{\textwidth}
                    \centering
                     \includegraphics[width=0.45\textwidth]{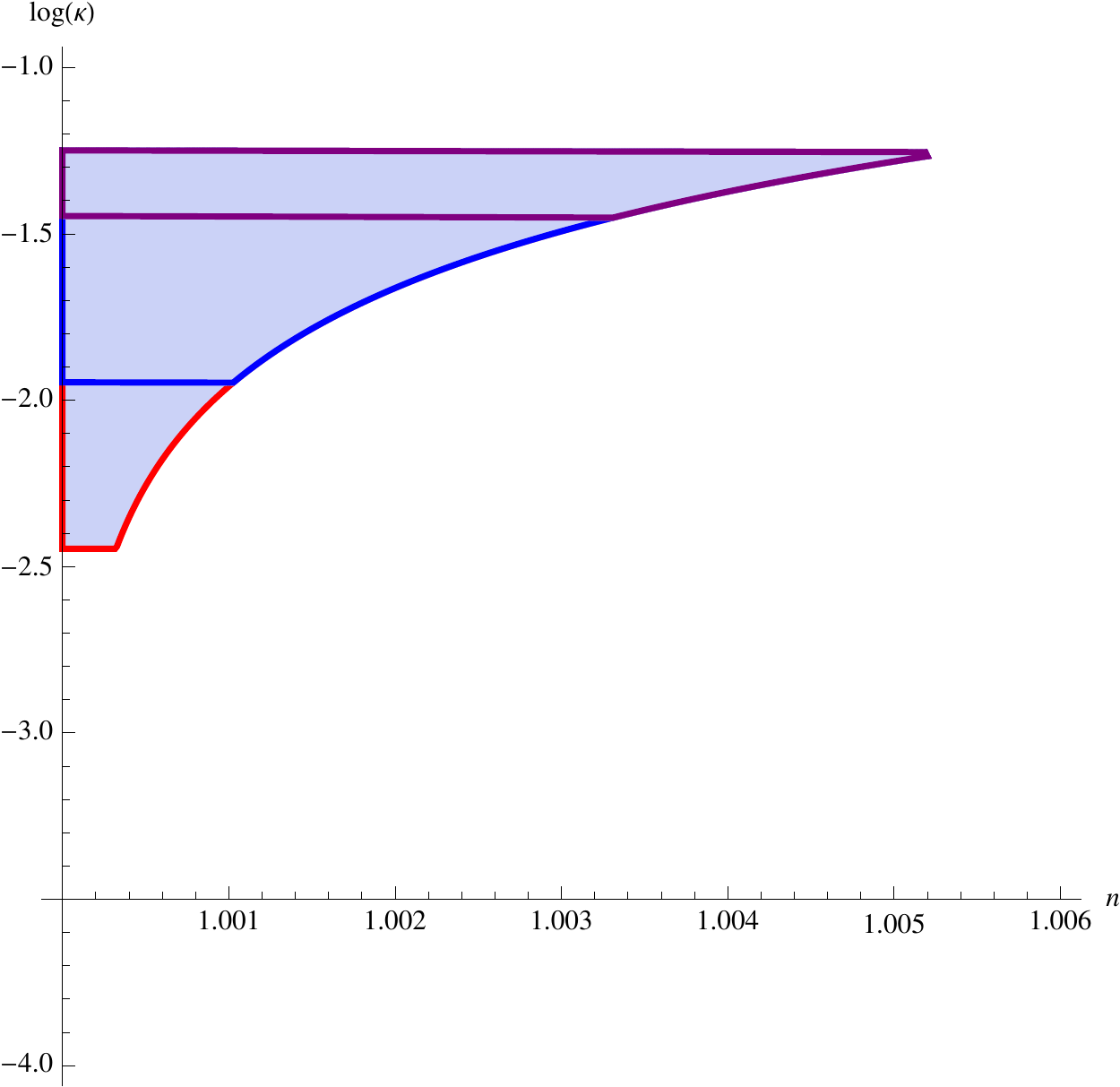}
                     \includegraphics[width=0.45\textwidth]{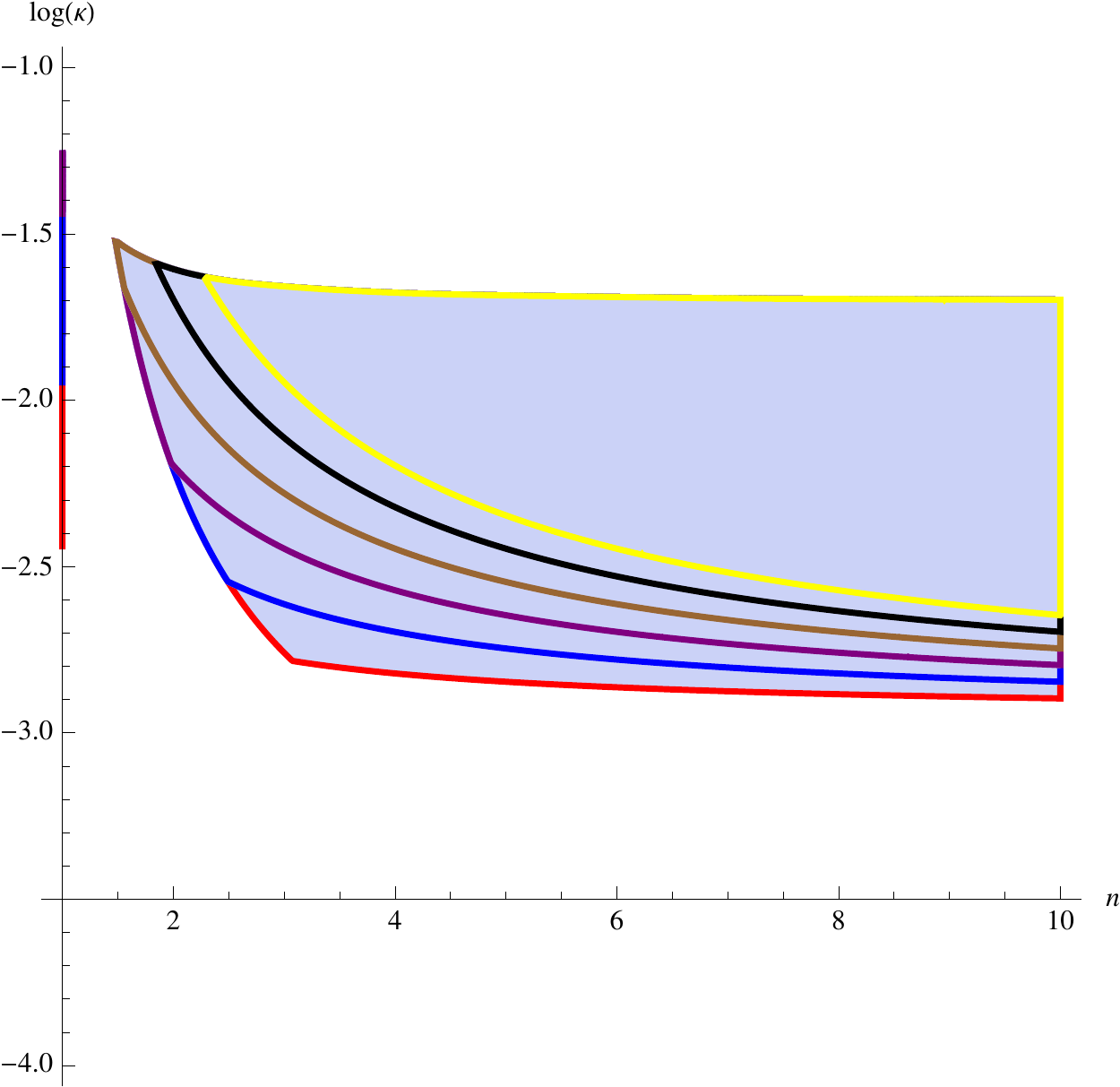}
                     \caption{One sigma.}
                     \label{fig:Lunar}
               \end{subfigure}
                \begin{subfigure}[b]{\textwidth}
                     \centering
                     \includegraphics[width=0.45\textwidth]{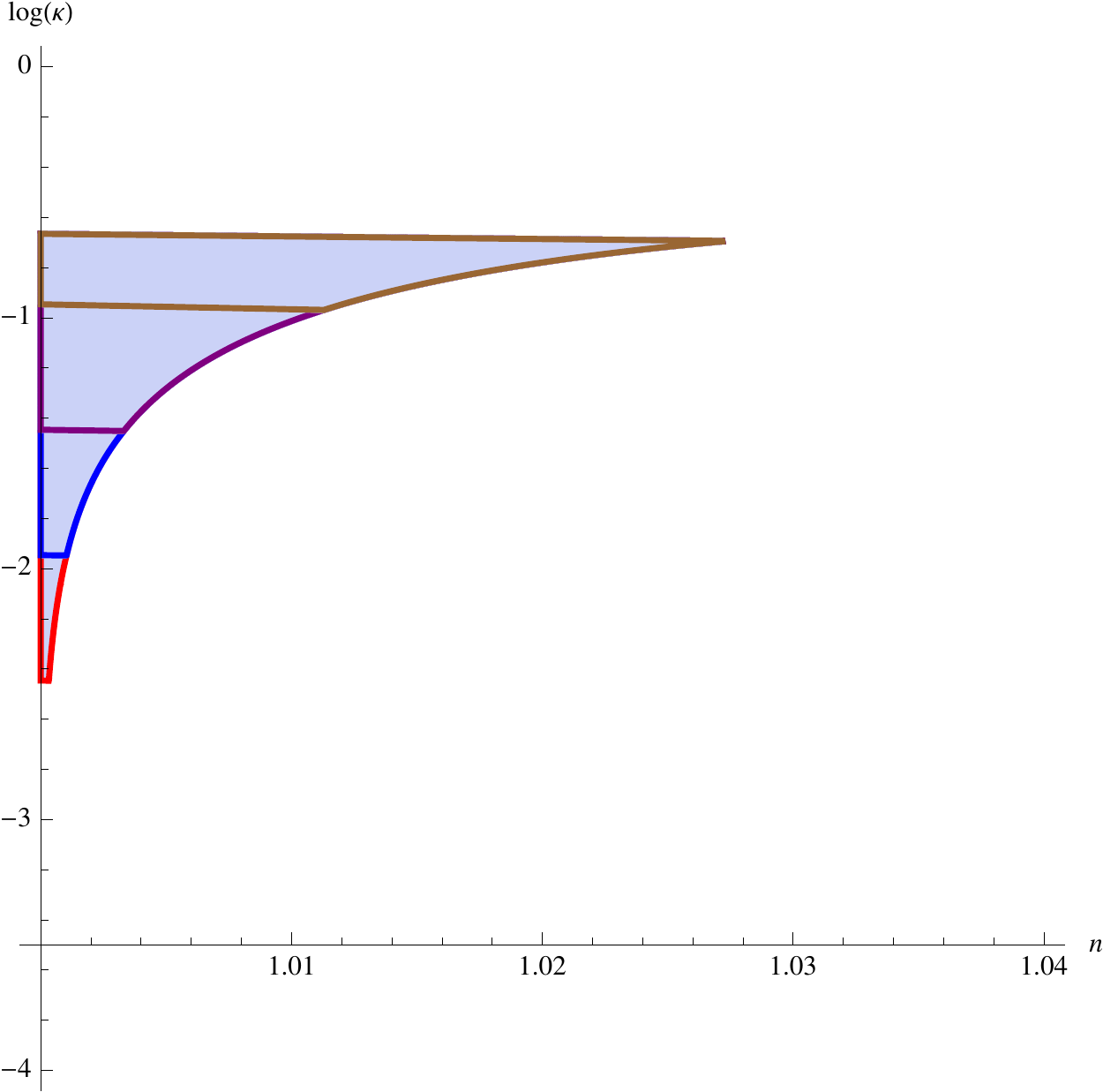}
                     \includegraphics[width=0.45\textwidth]{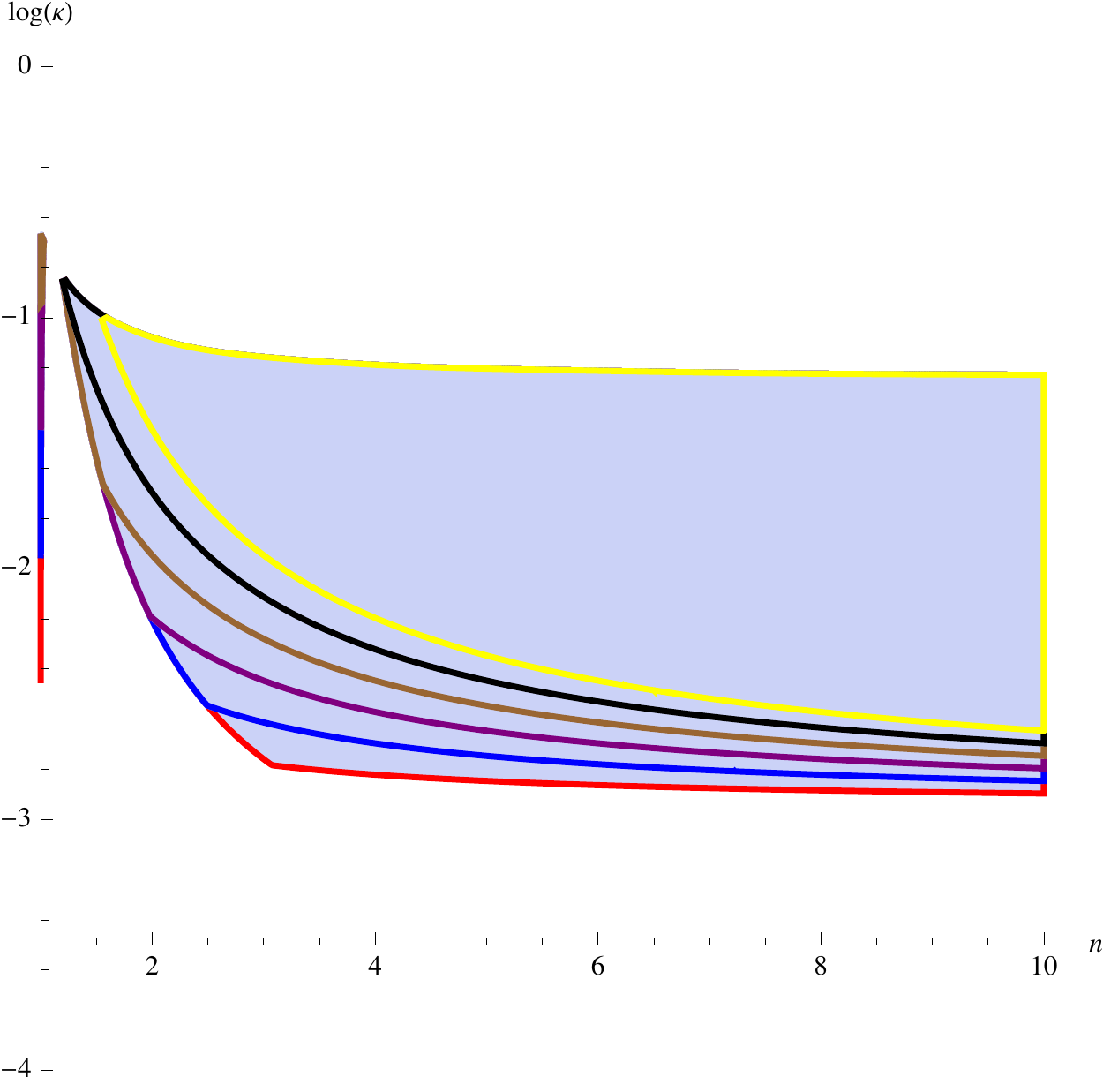}
                     \caption{Three sigmas.}
                                          \label{fig:Lunar2}
                \end{subfigure}
                \caption{The  values of $n$ and $\kappa$ which are consistent with the Lunar system data for $p=1,2,\cdots,6$ and the weak gravitational lensing at one and three sigmas . The smaller the regime is the larger $p$ is.}
\end{figure}
This lower bound allows us to neglects the corrections of the theory to the light speed travel time between the Moon and Earth. Notice that when \eqref{cDeltaTLunar} is not met, one should repeat the calculation of the Earth GM mass from raw data within our theory. This is clearly  computationally extensive. We have escaped from this task by considering the set of parameter that holds \eqref{cDeltaTLunar}. We further observe that the lower bound in \eqref{lunarExpansionAllowed} for all $p>0$ is larger than the lower bound in  \eqref{cDeltaTLunar} for $n\ge 1$.  The consistency with the Lunar system, then, requires \eqref{LunarConstraint}.
Utilizing \eqref{DPhiFastLeading} then simplifies \eqref{LunarConstraint} to
\begin{subequations}
\label{k3lunar}
\begin{eqnarray}
\kappa_3(n)&\equiv& \left(\frac{(2^{2n}-4) 2^n}{n}\right)^{\frac{1}{2n-1}} 10^{-4+\frac{4}{2n-1}}\,,\\
\kappa_3(n) &<&\kappa\,,
\end{eqnarray}
\end{subequations}
Notice that the lunar system provides the lower bounds on $\kappa$ while the gravitational lensing provide an upper bond.    The fig. \ref{fig:Lunar} shows the values of $n$ and $\kappa$ that are consistent with  \eqref{LunarConstraint} and \eqref{lunarExpansionAllowed} and one sigma level constraint in fig. \ref{knplotWGL}. The allowed regimes are disconnected.  The arbitrary large values of $n$ are allowed for $\kappa < 0.02$.
The fig. \ref{fig:Lunar2} depicts the values of $n$ and $\kappa$ that are consistent with  \eqref{LunarConstraint} and \eqref{lunarExpansionAllowed} and the three sigma level constraint of fig. \ref{knplotWGL}.

\section{On the consistency of the static spherical solutions}
\label{ConsistencySection}
At the end of section \ref{dynamics}, we have shown that the energy momentum tensor of the gauge fields and scalar vanish in the limit of the Einstein-Hilbert gravity. In this section we would like to compute the correction to the energy momentum tensor as we move away from the Einstein-Hilbert regime. In so doing, we notice that the following taylor expansions exist around the Einstein-Hilbert regime:
\begin{eqnarray}
\mu_n(x) &=& 1 - \frac{1}{n} x^{-n} \epsilon^2+ \frac{1+n}{2n^2} x^{-2n} \epsilon^4
-\frac{1+3n+2n^2}{6n^3} x^{-3n}\epsilon^6
+\cdots\,,\\
{\cal L}_n(x^2) &=& x \left(1- \frac{2}{(2-n)n} x^{-n} \epsilon^2+ \frac{1+n}{2(1-n)n^2} x^{-2n}\epsilon^4
-\frac{1+3n+2n^2}{3n^3(2-3n)} x^{-3n}\epsilon^6
+\cdots \right)\,,
\end{eqnarray}
wherein the auxiliary parameter of $\epsilon$ is turned on to systematically track the perturbation series.  At the end of the perturbative computation we set $\epsilon=1$. We also get:
\begin{eqnarray}
16 \pi G (T_{A_1 00} + T_{A_2 00})  &=& \frac{c^4}{k l^2} \left(\frac{k\tilde{r}}{r}\right)^{4-2n}
\left(
 \frac{2^{7-3n}}{n(n-2)} \epsilon^2
 -\frac{2^{5-6n}(3n-1)}{n^2(n-1)}\left(\frac{k\tilde{r}}{r}\right)^{-2n} \epsilon^4
 \right)\\
16 \pi G (T_{\phi_1 00} + T_{\phi_2 00}) &=&   \frac{c^4}{k l^2}
\left(\frac{k\tilde{r}}{r}\right)^{4-2n}
\left(
 -\frac{2^{5-n}}{n(n-2)} \epsilon^2
 +\frac{2^{3-2n}(3n-1)}{n^2(n-1)}\left(\frac{k\tilde{r}}{r}\right)^{-2n} \epsilon^4
 \right)\nonumber\\
\end{eqnarray}
So
\begin{eqnarray}
16 \pi G T_{00} \,=\,\frac{c^4}{k l^2} \left(\frac{k \tilde{r}}{r}\right)^{4-2n}
\left(-\frac{2^{5-3n}(4^n-4)}{(n-2)n} \epsilon^2
+ \frac{(4^{2n}-4)(3n-1)}{8^{2n-1}(n-1)n^2} \left(\frac{k \tilde{r}}{r}\right)^{-2n} \epsilon^4
\right)\nonumber\\
\end{eqnarray}
Let it be defined:
\begin{equation}
\bar{M}_{eff}(r) \,=\, \frac{1}{c^2} \int_0^r T_{00}(r') \, 4\pi r'^2 dr'
\end{equation}
$\bar{M}_{eff}(r)$ is the effective mass that  $T_{00}$ effectively induces in a sphere of radius $r$ around the origin. We notice that  only for
\begin{equation}
\frac{1}{2} < n\,,
\end{equation}
the effective mass is bounded. We further notice that for
\begin{equation}
 \frac{1}{2} \leq n < 1  \,,
\end{equation}
 and
\begin{equation}
 2 < n\,  ,
\end{equation}
the effective mass around the Einstein-Hilbert regime is negative while it is positive for
\begin{equation}
  1 \leq n\, \leq 2 .
\end{equation}
In the MOND regime we have the following perturbative series:
\begin{eqnarray}
\mu_n(x) &=& \sqrt{x}(1-\frac{1}{n} x^{\frac{n}{2}} \epsilon^2)\,,\\
{\cal L}_{n}(x) &=& \frac{2}{3} x^{\frac{3}{2}} (1- \frac{3}{n(3+n)}x^{\frac{n}{2}} \epsilon^2)\,,
\end{eqnarray}
The perturbation results to
\begin{equation}
16 \pi G T_{00}  = \frac{48 c^4}{ k l^2} \left(\frac{k\tilde{r}}{r}\right)^4 - \frac{32 \sqrt{2} c^4}{3 k l^2} \left(\frac{k\tilde{r}}{r}\right)^3 \left(1+ 3 \times 2^{\frac{n}{2}-1} \frac{(2^{n+1}-1) }{n(3+n)} \left(\frac{k\tilde{r}}{r}\right)^n \epsilon^2 \right)
\end{equation}
We observe that for all values of $n>0$, the tree-level $T_{00}$ is negative in the MOND regime for:
\begin{equation}
\frac{9}{2\sqrt{2}} \kappa < \frac{r}{\tilde{r}}\,.
\end{equation}
Now let it be defined
\begin{equation}
\tilde{M}_{eff}(r) \,=\, \frac{1}{c^2} \int_r^{L} T_{00}(r') \, 4\pi r'^2 dr'\,,
\end{equation}
$\tilde{M}_{eff}(r)$ is the effective mass that $T_{00}$ effectively induces between a sphere of a radius $r$ and a sphere of radius $L$. For a sufficient large value of $L$ we get:
\begin{equation}
\tilde{M}_{eff} \,=\,-\frac{\sqrt{2}}{3} \frac{c^2 \kappa^2}{G l^2} (2\tilde{r})^3 \ln(\frac{L}{k\tilde{r}})\,,
\end{equation}
For $|\tilde{M}_{eff}|$ at the order of $M$ then the perturbative solution around the flat geometry is not valid. This means that there is a minimum value for $L$ for which the perturbation is valid. We can find its order by demanding $\tilde{M}_{eff}=-M$:
\begin{eqnarray}
M_{0} & \equiv &  \frac{c^4}{a_0 G}= 1.21 \times 10^{54} kg,\\
\label{LConsistency}
\frac{L}{2\kappa } & =& \sqrt{\frac{GM}{a_0}} \exp(\frac{3}{\sqrt{2}}\sqrt{\frac{M_{0}}{M}}),
\end{eqnarray}
The finite value for $L$ means that the MONDian $\frac{1}{r}$ behavior  does not extend to infinity. At order $L$, the gravitational field must declines faster than $\frac{1}{r}$.  For the Sun, L reads
\begin{equation}
\frac{L_\odot}{\kappa} = 3.3 \times  10^{71658595155} m\,.
\end{equation}
For a galaxy of mass $M_g=10^{12} M_\odot$, we get
\begin{equation}
\frac{L_g}{\kappa } = 4.5 \times  10^{720216} m\,.
\end{equation}
For a clauster galaxy of mass $M_c=10^{16} M_\odot$, it yields
\begin{equation}
\frac{L_c}{\kappa } = 4.9 \times  10^{7222} m\,.
\end{equation}
Due to the size of the Universe we can not observe the deviation from $\frac{1}{r}$ in the deep MONDian regime.
\section{Cosmological Constant}
\label{CosmologiucalConstant}
Assuming that the constant of integration is not vanishing in  \eqref{ln} leads to the following term in the action:
\begin{equation}
S_{c_1}= -\frac{1}{16 \pi G l^2} \int \sqrt{-\det g } (\frac{1}{\tilde{k}_2}+\frac{1}{k_2}) c_1
\end{equation}
which is equivalent to adding a cosmological term to the action:
\begin{equation}\label{LambdaTheory}
\Lambda= \frac{3c_1}{\kappa l^2}
\end{equation}
\begin{figure}[tbp]
               \centering
                     \includegraphics[width=0.7\textwidth]{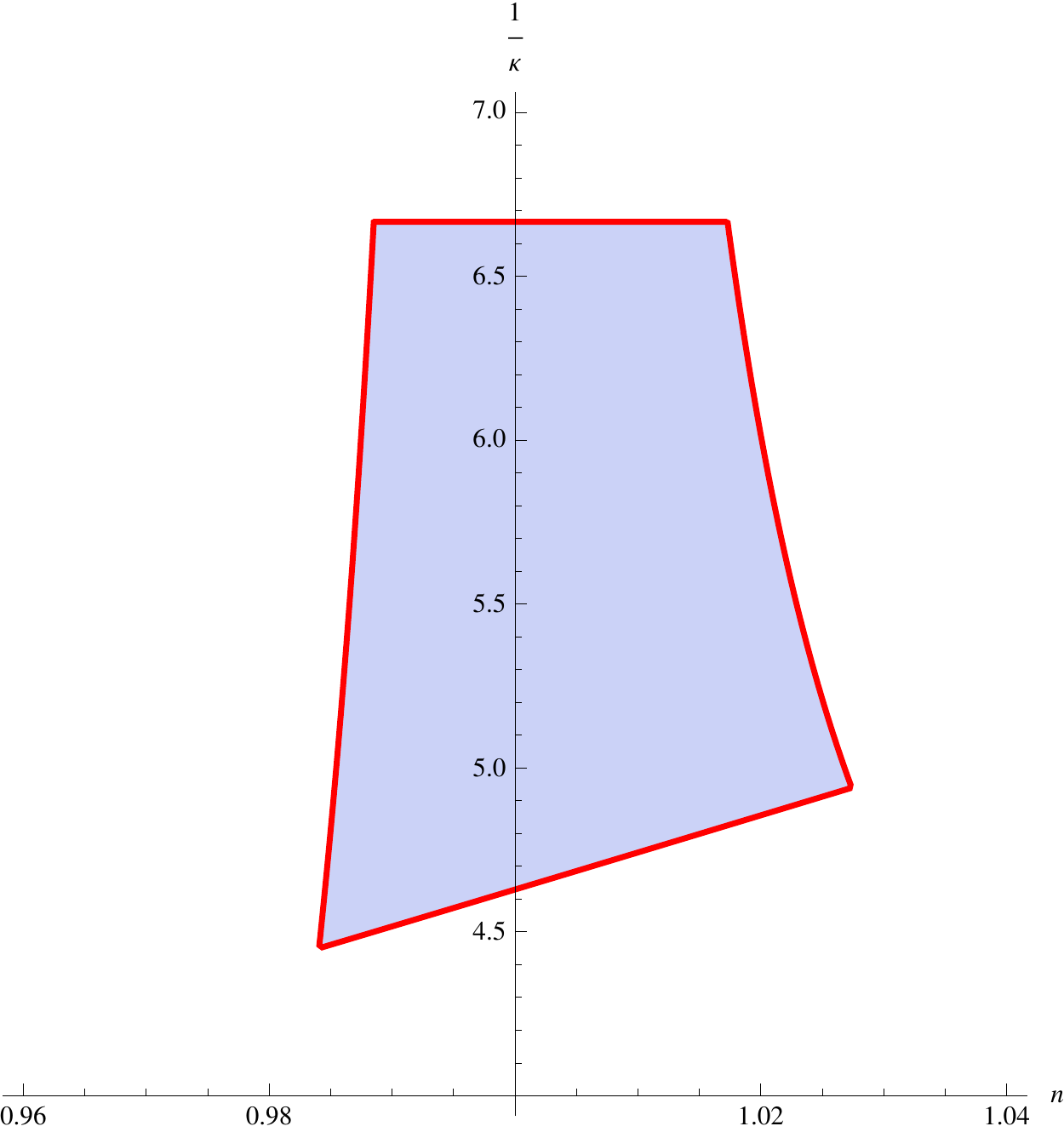}
                \caption{The  values of $\kappa$ and $n$ that are consistent with 3$\sigma$ weak gravitational lensing, Lunar system and \eqref{LambdaKappaConstraint}.}
                \label{kncosmology}
\end{figure}
The cosmological constant in term of $h$ and $\Omega_\Lambda$ reads
\begin{equation}
\Lambda \, =\, \frac{3}{c^2}\, H_0^2 \Omega_\Lambda\,.
\end{equation}
Table IV of  \cite{Tegmark:2003ud} reports  values of $\Omega_\Lambda$  and $H_0$ due  to +SN Ia observation as :
\begin{eqnarray}
\Omega_\Lambda &=& 0.725^{+0.039}_{-0.044}\,,\\
h &=& 0.599^{+0.090}_{-0.062}\,,
\end{eqnarray}
where $H_0 = 100\, h\, km\, s^{-1} MPc^{-1} = 3.24 \times 10 ^{-18} h s^{-1}$. Using these values, one finds
\begin{equation}\label{cosmologiacl_measured}
\Lambda_{\text{measured}} \,=\, 1.206^{+0.064}_{-0.073} \times 10^{-52} \frac{1}{\text{meters}^2} \,.
\end{equation}
Equating \eqref{cosmologiacl_measured} to \eqref{LambdaTheory} identifies $c_1$:
\begin{equation}
c_1 = \frac{\kappa^3}{6} \frac{c^4 \Lambda_{\text{measured}} }{a_0^2} = (18.70\pm 8.08) \,\kappa^3
\end{equation}
We demand $c_1$ and $\kappa$ to be at order one. This leads to
\begin{equation}
\label{LambdaKappaConstraint}
0.15 < \kappa <  0.70\,.
\end{equation}
Fig \ref{kncosmology} the allowed values of $n$ and $\kappa$ consistent with \eqref{LambdaKappaConstraint} and fig. \ref{fig:Lunar2}. This demands that $n=1$.
We prefer a value for $\kappa$ equals to the inverse of a natural number:
\begin{subequations}
\label{knLambda}
\begin{eqnarray}
\kappa&=&\frac{1}{5}\,,\\
n&=&1\,,\\
\frac{1}{c_1} &=& \,5 \cdots 11
\end{eqnarray}
\end{subequations}
and
\begin{subequations}
\label{knLambda2}
\begin{eqnarray}
\kappa&=&\frac{1}{6}\,,\\
n&=&1\,,\\
\frac{1}{c_1} &=& \,8 \cdots 20
\end{eqnarray}
\end{subequations}
reproduce the observed value of the cosmological constant, generates MOND and the weak gravitational lensing attributed to dark matter at three sigma, and are consistent with the lunar system data.
\begin{figure}[tbp]
               \centering
                     \includegraphics[width=0.7\textwidth]{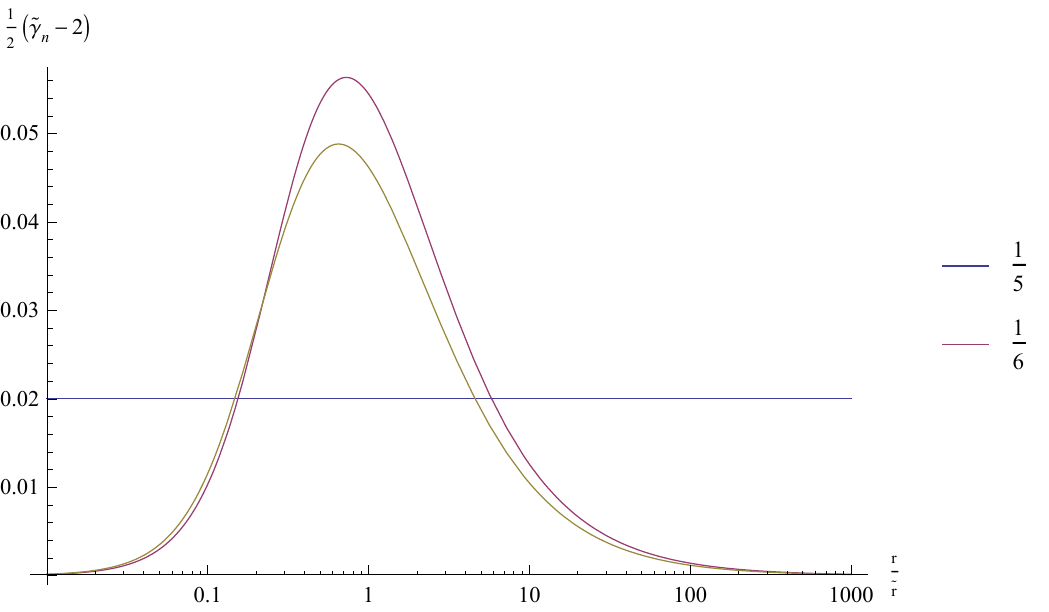}
                \caption{$\tilde{\gamma}_n(r)$ for $n=1$ and $\kappa=\frac{1}{5}$ and $\frac{1}{6}$.}
                \label{k156}
\end{figure}
 Fig.
\ref{k156} presents $\tilde{\gamma}$ for \eqref{knLambda}. We notice that the average of  $\Delta\gamma$ in    $\frac{r}{\tilde{r}}\in (0,10)$ for $\kappa=\frac{1}{5}$ reads:
\begin{equation}
<\frac{\bar{\gamma}-2}{2}> = \frac{1}{\frac{4 \pi}{3} (10 \tilde{r})^3 } \int_0^{10 \tilde{r}}  dr 4 \pi r^2 \frac{\tilde{\gamma}-2}{2}=0.014
\end{equation}
while for $\kappa=\frac{1}{6}$ it reads
\begin{equation}
<\frac{\bar{\gamma}-2}{2}> = \frac{1}{\frac{4 \pi}{3} (10 \tilde{r})^3 } \int_0^{10 \tilde{r}}  dr 4 \pi r^2 \frac{\tilde{\gamma}-2}{2}=0.017 \,.
\end{equation}
This means that these values of $\kappa$ are in fact in agreement with  \cite{Amendola:2013qna} that reports the average of $\gamma$.  We observe that only for $\frac{r}{\tilde{r}}\in (0.15,6)$, $\Delta\gamma=\frac{\tilde{\gamma}-2}{2}$ exceeds $0.02$. In particular the average of $\Delta\gamma$ in $\frac{r}{\tilde{r}}\in (\frac{1}{2},2)$ exceeds well beyond $0.02$:
 \begin{eqnarray}
k=\frac{1}{5} &\to& <\Delta\gamma> = 0.05 \,,\\
 k=\frac{1}{6} &\to& <\Delta\gamma> = 0.04\,.
 \end{eqnarray}
 This means that we should find the average of $<\Delta\gamma>$ for $\frac{r}{\tilde{r}}\in (\frac{1}{2},2)$ rather than its average over whole of the space.  This is a prediction of the theory. This prediction can be attested by refining the current data or specifically designed future observations.

\section{Cosmological Solutions}
\label{Cosmology}
We accept that cosmology is described by a homogeneous isotropic solution in the sense that the the large matter inhomogeneities in small scales do not produce significant effects in large scales \cite{Green:2014aga}.

The field content of a homogeneous isotropic time dependent solution of the theory is represented by
\begin{eqnarray}
ds^2 &=& -dt^2 + a(t)^2 d|\vec{x}|^2\,,\\
\phi_1 &=& \phi_1(t)\,,\\
\phi_2 &=& \phi_2(t)\,,\\
\label{VGF}
A_\mu^1 &=& A_\mu^2 = (0,0,0,0)\,.
\end{eqnarray}
Notice that the gauge fields vanish due to symmetries.  We represent the local matter density by $\rho=\rho(t)$, and its four velocity by:
\begin{equation}
u_\mu = (u_0,0,0,0)\,,
\end{equation}
\eqref{umuconst} implies that
\begin{equation}
e^{2\bar{\phi}} u_0^2 = 1 \to u_0= e^{-\bar{\phi}}\,,
\end{equation}
So the matter current reads:
\begin{equation}
\label{CosmologyJ}
J_\mu = \rho e^{-\bar{\phi}} (1,0,0,0)\,.
\end{equation}
We observe that a vanishing gauge field does not satisfy \eqref{gaugeequations}.\footnote{This paradox also exists in ordinary QED: The electric field of a uniformly charged space vanish due to symmetries but the vanishing field is not consistent with the gauss law.} We need to change our theory to become consistent with cosmology. To do so define:
\begin{equation}
\bar{J}_\mu = \frac{\int d^3x \sqrt{-\det g} J^\mu}{\int d^3x \sqrt{-\det g}}\,,
\end{equation}
Next alter the matter action \eqref{MatterAction} to:
\begin{equation}
S_M = \int d^4x \sqrt{-\det g} (-\frac{1}{2} \rho  e^{2 \bar{\phi} } g_{\mu\nu} u^\mu u^\nu - A_\mu (J^\mu-\bar{J}^\mu))
\end{equation}
 The equations of motion  of the gauge-fields then follow
\begin{subequations}
\label{Cosmologygaugeequations}
\begin{eqnarray}
\nabla_\nu F^{\nu\mu}_1 & =& 16 \pi \kappa_1 G (J^\mu-\bar{J}^\mu)\,,\\
\label{Cosmologygaugeequationsb}
\nabla_\nu \left({\cal L}'(-\frac{l^2}{4 }F^2_{\alpha\beta} F_{2}^{\alpha\beta})F^{\nu\mu}_2\right) & =& 16 \pi \kappa_2 G (J^\mu-\bar{J}^\mu)\,.
\end{eqnarray}
\end{subequations}
 Since  any  local mass distribution in a non-compact universe holds  $\bar{J}_\mu=0$ then  the solutions of the previous sections are left intact. The homogeneous isotropic cosmological solution \eqref{CosmologyJ}, however, holds:
\begin{equation}
J^\mu-\bar{J}^\mu \,=\, 0\,.
\end{equation}
The equations of motion of the gauge-fields \eqref{Cosmologygaugeequations}, therefore, are in agreement with a vanishing gauge fields \eqref{VGF}.

In order to address the dynamics of $\phi_2$, we need to identify ${\cal L}(x)$ for negative value of $x$ because
\begin{eqnarray}
-\partial_\mu \phi_2 \partial^\mu \phi_2 &=& - \phi_2'(t)^2 < 0
\end{eqnarray}
 We consider the simplest choice:
\begin{equation}
{\cal L}(x)\,=\, x  + c_1  \qquad ~ \forall x<0\,.
\end{equation}
Notice that $c_1$ should be present in order to have $\cal L$ in $C^1$. The equations of motion of the scalars then read:
\begin{subequations}
\label{CosmologyScalarsEquations}
\begin{eqnarray}
\Box \phi_1 &=& 16 \pi \tilde{\kappa}_1 G \rho \,,\\
\label{CosmologyScalarsEquationsb}
\Box \phi_2 &=& 16 \pi \tilde{\kappa}_2 G \rho \,,
\end{eqnarray}
\end{subequations}
Due to the form of these equations, \eqref{ConsCoupling} and choosing appropriate boundary conditions we get
\begin{equation}
\bar{\phi}=\phi_1 + \phi_2 = 0 \,.
\end{equation}
This means that the scalars as well as the gauge fields do not affects the trajectories of the particles. The net contribution of the energy momentum tensors of the scalar and gauge fields vanish too:
 \begin{eqnarray}
T_{\phi_1\mu\nu} + T_{\phi_2\mu\nu} & =& 0 \,, \\
T_{A_1\mu\nu} + T_{A_2\mu\nu} & =& 0 \,.
\end{eqnarray}
Notice that the $c_1$ in the definition of $\cal L$ generates the cosmological constant. The equation of motion of the metric then simplifies to:
\begin{eqnarray}
G_{\mu\nu} + \Lambda g_{\mu\nu} = 8 \pi G T_{M\mu\nu}\,,
\end{eqnarray}
where $\Lambda$ is given in \eqref{LambdaTheory}. So the cosmological solution of the theory coincides to that of the Einstein-Hilbert action. The late-time behavior of the theory is in agreement with the observation.

\section{Conclusions and discussions}
\label{summary}
This research has been started by asking how precise the space-time geometry can be approximated by a pseudo Riemann geometry. Can quantum gravity or dark matter/energy be naturally resolved in geometries beyond Riemann? Can the degrees of freedom for dark fields be assigned to a deviation from Riemann geometry instead of being added to the table of the ordinary elementary particles? Can the MOND theory \cite{MOND} be naturally realised in a deviation from the Riemann geometry?

In order to answer some of these questions, we have considered a deviation from the Riemann geometry toward the Randers' geometry in the section \ref{RandersSection}. This deviation includes two gauge fields and two scalars in addition to the metric. It possesses four coupling constants, a length scale and a functional.

In the section \ref{dynamics}, we have fixed the dynamics of the gauge fields such that the strong field limit of the theory coincides to that of the Einstein-Hilbert theory. This has reduced the number of the free coupling constants  to two. We have further fixed the dynamics of the theory such that its weak field limit reproduces the AQUAL \cite{AQUAL} and MOND theories. This has mapped the length parameter of the theory to the critical acceleration of the MOND, and has fixed the asymptotic behaviours of the functional of theory.

In section \ref{staticsolutions}, we have presented the static solutions in the MOND and Newtonian regimes of the theory. We have studied the light trajectory around static solutions. We have calculated the $\gamma$ parameter in the MOND regime of theory. We have required the $\gamma$ parameter in the MOND regime to be one. This requirement has reduced the number of coupling constants to one.

We have considered a family for the functional of the theory in the section \ref{WeakGravitationalLensing}. This family can be expressed in terms of the Gauss's hypergeometric function and a constant of integration. We have assumed that the constant of the integration is sufficiently small and does not affect the space-time geometry at the considered vicinity of  the spherical mass distribution. We have ignored the negligible contribution of the gauge fields and scalars in the MOND regime to the metric by their contributions to the energy momentum tensor. In doing so, we have found the exact spherical solutions of the theory. We have calculated the $\gamma$ parameter in the whole of the space. We have reported that requiring the $\gamma$ parameter to be consistent with the constraint reported in   \cite{Amendola:2013qna} sets a strong upper bound on the coupling constant of the theory as depicted in the fig. \ref{knplotWGL}.  We have observed that this bound  is in agreement with the value of $\gamma$ measured at the surface of the Sun as depicted in fig. \ref{fig:SunGamma}.

We have studied the Lunar System constraints on the theory in the section \ref{LunarSystemConstraints}. We have used the Laser Lunar Ranging measurements \cite{LLR} and LAGEOS \cite{LAGEOS} data. We have  assumed that the space-time geometry around the earth provides a perturbation to  that predicted by the Einstein-Hilbert gravity. We have sufficed to the inclusion of  the leading perturbation and assumed that this perturbation is smaller by the factor of $10^{-p}$ from the Newtonian gravity. We have identified the regime of the parameters wherein the theory does not affect the time of the travel of the signals from the Earth to the Moon with the precision reported in \cite{LLR}. We have shown that these considerations and \cite{LLR,LAGEOS}  provide a lower bound on the coupling constant of the theory as depicted in  the fig. \ref{fig:Lunar} and \ref{fig:Lunar2}.

We have studied the consistency of our approach in \ref{ConsistencySection}. We have shown that as we move toward the asymptotic infinity, the neglected contribution of the energy-momentum tensor of the gauge and scalar fields accumulates. For any given mass, we have calculated the length scale wherein the energy-momentum tensor of the gauge and scalar fields must be considered, the result has been expressed in \eqref{LConsistency}. We have reported that for any given astronomical mass, this length scale is far larger than the size of the observed Universe. Notice that  the kinetic term of one of the gauge fields and one of the scalars have negative sign and they might cause a new difficulty in the quantisation of the geometry that here is composed of the metric, two scalars (non-minimally coupled to matter) and two gauge fields.  We have shown that the  strong field limit of theory  coincides to the Einstein-Hilbert gravity also at the level of quantum field theory. We have studied  the classical solutions (IR limit) and we have shown that the theory is classically consistent in its IR limit.  The theory is not trivial. Addressing the quantum stability of the theory in the interpolating IR-UV regime requires addressing  quantisation of the geometry and is not within the scope of this work. 

 We have mapped the constant of the integration in the definition of the functional of the theory to  the observed cosmological constant in the section of \ref{CosmologiucalConstant}. This map is natural in the sense that it needs a tuning at order one. So we have shown that the theory naturally accommodates the cosmological constant. We further have shown that the accommodation  identifies the only free coupling constant of the theory.  We have reported that the  identified theory predicts that the $\gamma$ parameter in the interpolating regime deviates from 1 by few percent albeit only in a specific section of the interpolating regime. This prediction can be attested by refining the current data or specified future observations.

We have studied the cosmology of the theory in section \ref{Cosmology}. We have slightly changed the coupling of the gauge fields to the matter current in order to have a homogeneous isotropic time-dependent solution. This change leaves intact the static solutions of the previous sections. We have reported that the isotropic homogeneous solution of the theory coincides to that of the Einstein-Hilbert theory with a cosmological constant. So the late time cosmology of the theory is indeed in accord with observation. We have left the study of the early universe in particular that of the fluctuations of CMB in the theory to a future study.

In summery, we have presented a deviation of the physical length  from the Riemann geometry toward the Randers' geometry. We have constructed  a consistent second-order  relativistic theory of gravity that dynamically reduces to the Einstein-Hilbert theory for strong and Newtonian gravity while its weak gravitational regime reproduces MOND and the gravitational lensing attributed to the dark matter halo. It also naturally accommodates  the observed value of the cosmological constant. We have reported that it
predicts a small deviation for the post Newtonian parameter $\gamma$ from $1$ in a specific  regime that interpolates the Newtonian regime to the MOND regime.  The deviation  is consistent with the reported observations but can possibly be detected by fine-tuned refinements of the current or future data. We, however, do not jump into the conclusion that all the reported aspects of dark matter, including the acoustic peaks in CMB or the Bullet cluster, can be reproduced by this model. Since the model is mathematically simple and phenomenologically adaptable, we do not refute the possibility that further aspects of dark matter can be accommodated in degrees of freedom of the space-time geometry that encodes some deviations from the Riemann geometry. Future works will investigate  this possibility.

\section{Aknowledgments}
This work was  supported by a grant from the National Elites Foundation (NEF) of Iran, and a grant from IPM.  I  thank the K. N. Toosi University of Technology  that laid me off from my assistant professor position at the physics department because of my high pitched voice, and froze my NEF grant.\footnote{\href{http://www.theguardian.com/world/iran-blog/2015/mar/16/iranian-professor-qasem-exirifard-loses-job-due-to-feminine-voice}{http://www.theguardian.com/world/iran-blog/2015/mar/16/iranian-professor-qasem-exirifard-loses-job-due-to-feminine-voice}}
\providecommand{\href}[2]{#2}\begingroup\raggedright

\end{document}